\documentclass[reprint,superscriptaddress,amsmath,amssymb,aps,prx,floatfix]{revtex4-2}

\usepackage{graphicx}
\usepackage{dcolumn}
\usepackage{bm}
\usepackage{tikz}
\usetikzlibrary{calc}
\usepackage{placeins}
\usepackage{wrapfig}
\usetikzlibrary{positioning, backgrounds, calc}
\usetikzlibrary{arrows.meta,positioning,fit,backgrounds}

\newcommand{\PGrevision}[1]{#1}

\begin{document}

\preprint{APS/123-QED}

\title{Information bottleneck for learning the phase space of dynamics from high-dimensional \PGrevision{observations}}

\author{K.~Michael Martini}
\thanks{Equal Contribution}
\affiliation{Department of Physics, Emory University, Atlanta, GA, USA}
\affiliation{Initiative in Theory and Modeling of Living Systems, Emory University, Atlanta, GA, USA}

\author{Eslam Abdelaleem}
\thanks{Equal Contribution}
\affiliation{Department of Physics, Emory University, Atlanta, GA, USA}
\affiliation{Schools of Physics and Psychology, Georgia Institute of Technology, Atlanta, GA, USA}

\author{Paarth Gulati}
\thanks{Equal Contribution}
\affiliation{Department of Physics, Emory University, Atlanta, GA, USA}
\affiliation{Initiative in Theory and Modeling of Living Systems, Emory University, Atlanta, GA, USA}

\author{Ilya Nemenman}
\thanks{Corresponding author: \texttt{ilya.nemenman@emory.edu}}
\affiliation{Department of Physics, Emory University, Atlanta, GA, USA}
\affiliation{Initiative in Theory and Modeling of Living Systems, Emory University, Atlanta, GA, USA}
\affiliation{Department of Biology, Emory University, Atlanta, GA, USA}

\begin{abstract}
Identifying the dynamical state variables of a system from high-dimensional observations is a central problem across physical sciences. The challenge is that the state variables are not directly observable and must be inferred from raw high-dimensional data without supervision. Here we introduce DySIB (\textbf{Dy}namical \textbf{S}ymmetric \textbf{I}nformation \textbf{B}ottleneck) as a method to learn low-dimensional representations of time-series data by maximizing predictive mutual information between past and future observation windows while penalizing representation complexity. This objective operates entirely in latent space and avoids reconstruction of the observations. We apply DySIB to 
\PGrevision{simulated videos of a rigid-rod pendulum}
where the underlying state space is known. The method, with hyperparameters of the learning architecture set self-consistently by the data, recovers a two-dimensional representation that matches the \PGrevision{dimensionality and  topology}
of the pendulum phase space, with the learned coordinates aligning smoothly with the canonical angle and angular velocity. These results demonstrate, 
\PGrevision{for this controlled system,}
that predictive information in latent space can be used to recover interpretable dynamical coordinates directly from high-dimensional data.
\end{abstract}
\maketitle

\section{Introduction}
\label{sec:intro}
Many natural and synthetic dynamical systems are governed by a small number of state variables. For example, the state of a physical pendulum is fully specified by its angle and angular velocity; a neural action potential can be modeled by the membrane voltage and the gating states of just a few ion channels~\cite{hodgkin1952quantitative}; collective behavior in inanimate matter and in animal flocks is organized by a small number of slowly varying collective modes~\cite{toner1995long,goldenfeld2018lectures,cavagna2023natural}. The dynamics remain low-dimensional even when observed through high-dimensional, partially redundant measurements, such as the sequence of video images formed on our retina or on a CCD camera when watching the pendulum or the flock.

Detecting these low-dimensional variables in high-dimensional data, and understanding how they govern each other's dynamics, is one of the central problems of theoretical physics. The field has developed a powerful set of tools for this task, exploiting symmetries, locality, separation of scales, and the resulting hierarchy of effective theories. The Landau program, in particular, identifies the slow collective modes of a system as order parameters derived from broken symmetries, and writes equations of motion directly in terms of those modes. This approach, however, requires locality, symmetries and conservation laws known {\em a priori}, often augmented by physical intuition about which scales matter. For systems where these symmetries and locality do not exist or are not known, such as gene-regulatory networks, animal behavior, or raw video of physical experiments, a data-driven approach to identify the effective variables is required. With the advent of modern AI, where the same low-dimensional variables that describe a system (known in the AI-speak as latent representations or embeddings) underpin nearly all applications, one can dream of an approach in which AI itself is used to ``learn new physics'' from raw data~\cite{daniels2019automated,bapst2020unveiling,schmitt2024machine,yu2025dusty}, including the detection of physically meaningful, low-dimensional, interpretable representations of the systems of interest~\cite{stephens2008dimensionality,cunningham2014dimensionality,schoenholz2016structural,noe2017collective,cubuk2017structure,pandarinath2018inferring,ahamed2021capturing,colen2021machine,supekar2023learning}.

Many AI approaches have indeed been tried. The best-known ones tackle a somewhat orthogonal problem: starting from approximately correctly identified low-dimensional state variables, they focus on detecting dynamical relations among them~\PGrevision{\cite{schmidt2009distilling,daniels2015automated,brunton2016discovering,mangan2019model,frishman2020learning,reinbold2021robust,gurevich2024spider,liu2021machine}}. A second large class of approaches uses autoencoders~\PGrevision{\cite{lusch2018deep,champion2019data,linot2020deep,page2021revealing,chen2022automated,vlachas2022multiscale,lu2020extracting}}, where a low-dimensional latent representation is enforced, while requiring that the original data can be reconstructed from it with minimal error. Then dynamical relations are sought between the latent representations of pasts and futures. However, the information needed to reconstruct the data does not have to coincide with the information relevant to the dynamics. Thus, such approaches often preserve all salient variation in their embeddings, including variability with little relation to the dynamics.

Modern AI has focused on prediction instead of auto-reconstruction, chiefly through autoregressive data-space generative modeling~\cite{brown2020language,lai2025panda}. Here, a latent representation is built to predict the next state of the system in the data space---for example, generating the next frame of the high-dimensional movie that captures the dynamics, or the next snapshot of a forecast weather field~\cite{lam2023learning}. In the linear setting, focusing on prediction rather than auto-reconstruction sometimes lowers the data requirements~\cite{abdelaleem2024simultaneous,swain2025better,mergny2025spectral,baharav2025stacked}, and one could hope for similar results in nonlinear cases as well, so that predicting the future may be easier than reconstructing the now. This pivot to time series prediction makes AI considerably more relevant to physical sciences, and yet the goals of generative AI remain fundamentally different from what physical modeling requires. Indeed, Newton's laws do not predict the next video frame but the future of the latent variables describing it, namely position and velocity of a particle in that frame. In other words, physical understanding lives in the latent space, and the dynamics it describes must operate there directly~\cite{wiskott2002slow,balestriero2025lejepa,maes2026leworldmodel}. Under some conditions, such latent-prediction models, even nonlinear ones, are provably more sample-efficient than data-space-based generative ones, since they need only enough data to sample the much lower-dimensional latent space~\cite{martini2024data,vanassel2025joint}. Nonetheless, adoption of latent-space predictive approaches for modeling dynamics in the AI community has been slow. Where they have been adopted successfully~\cite{oord2018representation,schmitt2023information,schmitt2024model,meng2024bayesian,balestriero2025lejepa,maes2026leworldmodel}, the focus has rarely been on {\em low-dimensional, physically interpretable} embeddings. \PGrevision{Notable exceptions include dynamical components analysis (DCA) and compressed predictive information coding (CPIC), which use predictive information to identify low-dimensional representations of time-series data~\cite{meng2022compressed,clark2019unsupervised}. Separately,} models emerging from the AI community often learn past-to-future maps that do not preserve the differential structure of physical dynamics, where the change in the state variables over a short time interval must remain small.

Overall, this suggests that the problem of using high-dimensional 
\PGrevision{observations}
(e.g., videos) to learn low-dimensional embeddings that are predictive of their own temporal evolution, and 
\PGrevision{can be interpreted}
as the phase-space coordinates of physical dynamical systems, remains unsolved. In this paper, we propose and validate a solution to this problem, building on a principled framework for constructing information-theoretic loss functions for AI applications, the Deep Variational Multivariate Information Bottleneck~\citep{abdelaleem2025deep}. Extending the well-known Information Bottleneck construction~\citep{tishby2000information,alemi2017deep}, we seek to compress the representation of the high-dimensional vector describing the current state of a dynamical system in such a way that small, nearly differential, changes to it make it maximally informative, and hence predictive, of the similarly (that is, symmetrically) compressed future state of the system. In other words, the entire prediction is forced to live in the latent representation space, as it typically does in physics. We call our approach DySIB (\textbf{Dy}namical \textbf{S}ymmetric \textbf{I}nformation \textbf{B}ottleneck).

To test our approach, we apply it to 
\PGrevision{low-resolution videos of a simulated pendulum}
from~\cite{chen2022automated}, where the known physics allows for an easy interpretation of the results.
\PGrevision{We choose a system with a two-dimensional phase space deliberately. There, the fixed points and separatrices help characterize the phase portrait up to a continuous change of coordinates, so the learned and true phase spaces can be compared directly, which is considerably harder in higher dimensions.}
DySIB, with all of its tunable parameters set self-consistently by the data, recovers a two-dimensional representation with a meaningful phase-space structure, easily interpretable as ``correct'' by anyone with mastery of introductory physics. The learning of this low-dimensional phase space is possible with small datasets and with small neural networks (NNs), demonstrating the power of encoding physical inductive biases into the AI architecture.

The rest of the paper is organized as follows. In Sec.~\ref{sec:method} we review the theory behind our approach and introduce DySIB and its implementation. In Sec.~\ref{sec:results} we present results on the pendulum dataset. Finally, in Sec.~\ref{sec:discussion} we discuss limitations of our approach, and speculate on how it can be applied to other systems to learn {\em new} physics, rather than to recreate the known one.

\begingroup
\let\MakeTextUppercase\relax
\section{DySIB: Dynamical Symmetric Information Bottleneck}
\label{sec:method}
\endgroup

\subsection{Background: Information Bottleneck and its symmetric variant}
\label{sec:ib_background}

The Information Bottleneck (IB)~\citep{tishby2000information} formalizes the search for a compressed latent representation that preserves what matters and discards what does not. Given a high-dimensional signal $X$ and a supervising (or the \emph{relevance}) variable $Y$, one seeks a representation of the signal, $Z$, that is informative about $Y$ while retaining as little information about $X$ as possible. This is achieved by minimizing the loss
\begin{equation}
    \mathcal{L}_\mathrm{IB} = I(X;Z) - \beta\, I(Z;Y),
    \label{eq:ib}
\end{equation}
where $I(A;B)$ denotes the mutual information between $A$ and $B$~\citep{cover2006elements}, and the trade-off parameter $\beta$ sets the balance between the compression and the preservation of relevant information. In the limit $\beta \to \infty$, $Z$ becomes a sufficient statistic of $X$ for $Y$. Since the data processing inequality implies $I(Z;Y)\leq I(Z;X)$, the construction provides a natural measure of how close the latent representation is to capturing {\em all} the relevant information in the signal.

Real-world data are often multimodal, with several observed variables $X_1, X_2, \ldots$ each carrying partly overlapping information about the system of interest. The Multivariate Information Bottleneck (MIB)~\citep{friedman2001multivariate} encodes this structure in two Bayesian networks. An encoder graph specifies the statistical dependencies by which latent variables compress the observations, and a decoder graph specifies which of those compressed representations the latent variables are required to preserve. Writing the multiinformation~\citep{studeny1998multiinformation} of each graph, $I_\mathrm{graph}$, as a sum of mutual informations between each child node in the graph and the union of all of its parent nodes, the MIB loss takes the form
\begin{equation}
    \mathcal{L}_\mathrm{MIB} = I_\mathrm{encoder} - \beta\, I_\mathrm{decoder},
    \label{eq:mib}
\end{equation}
which generalizes Eq.~\eqref{eq:ib} to arbitrary patterns of compression and relevance relations~\citep{friedman2001multivariate}.

The Symmetric Information Bottleneck (SIB) is the special case of MIB in which two observed variables $X$ and $Y$ are both high-dimensional, such as the past and the future of large-dimensional observations of a dynamical system, and must be compressed symmetrically, though not necessarily equally. The encoder graph maps $X \to Z_X$ and $Y \to Z_Y$ via two independent compressions, and the decoder graph requires $Z_X$ and $Z_Y$ to be maximally informative about each other, thus preserving the information in the {\em latent} space. The resulting loss is
\begin{equation}
    \mathcal{L}_\mathrm{SIB} = I(X;Z_X) + I(Y;Z_Y) - \beta\, I(Z_X;Z_Y).
    \label{eq:sib}
\end{equation}
Formal recursive solutions for the optimal $p(z|x)$ in IB, and for $p(z_x|x), p(z_y|y)$ in SIB, are known~\citep{tishby2000information,friedman2001multivariate}, but computing them in practice is hard and typically requires variational assumptions about the structure of these distributions~\citep{jordan1999introduction}. As a result, the IB family of approaches became useful in high-dimensional, real-world problems only with the advent of deep neural networks (DNNs).

\subsection{Background: Deep Variational SIB}
\label{sec:dvsib_background}

The Deep Variational Symmetric Information Bottleneck (DVSIB)~\citep{abdelaleem2025deep} is a DNN-based variational implementation of Eq.~\eqref{eq:sib}. The latent variables are obtained using variational encoders, $q_X(z_x|x)$ and $q_Y(z_y|y)$, each implemented by a DNN. Training optimizes the usual evidence lower bound (ELBO)~\citep{jordan1999introduction,kingma2013auto,alemi2017deep}, which is tight when the variational posteriors $q_X(z_x|x)$ and $q_Y(z_y|y)$ approximate the true posteriors $p(z_x|x)$ and $p(z_y|y)$ implied by the data, and the variational priors used for $z_x$ and $z_y$ approximate the corresponding latent marginals. We parametrize these variational encoders as Gaussians with trainable means $\mu_X(x), \mu_Y(y)$ and log-variances $\ell_X(x), \ell_Y(y)$ such that
\begin{equation}
\begin{aligned}
    q_X(z_x|x) &= \mathcal{N}(\mu_X(x), \mathrm{diag}(\exp(\ell_X(x)))), \\
    q_Y(z_y|y) &= \mathcal{N}(\mu_Y(y), \mathrm{diag}(\exp(\ell_Y(y)))).
\end{aligned}
\end{equation}
Whenever during the optimization or inference one needs to produce samples in the latent spaces, $z_x$ and $z_y$, these are then drawn from these variational posteriors using the usual reparameterization trick \citep{kingma2013auto}.

The DVSIB objective is then given by:
\begin{equation}
\label{eq:dvsib}
    \mathcal{L}_{\text{DVSIB}} = \tilde{I}^E(X; Z_X)  +\tilde{I}^E(Y; Z_Y) - \beta \tilde{I}^D(Z_X; Z_Y), 
\end{equation}
where the first two terms are variational upper bounds on the mutual information $I(X; Z_X)$ and $I(Y; Z_Y)$, and are implemented as KL divergences between the variational posteriors and standard normal priors (see App.~\ref{SI:variational} for details). The third term is a variational lower bound on the mutual information between $Z_X$ and $Z_Y$. The superscripts $E,D$ follow the DVSIB nomenclature distinguishing 
the encoder-side (compression) term(s) from the decoder-side (predictive) term(s), respectively. 

In practice, the decoder term can be implemented by many different mutual information estimators, as discussed in~\citep{abdelaleem2025accurate}. We use the InfoNCE estimator~\citep{oord2018representation}, which contrasts matched $(z_x, z_y)$ pairs against mismatched pairs. Throughout, no reconstruction of $X$ and $Y$ is used and the objective is defined entirely in terms of the encoders and the latent variables, so that the entire training and subsequent inference stays in the latent space, paralleling JEPA approaches~\citep{balestriero2025lejepa,maes2026leworldmodel}. 

The trade-off parameter $\beta$ controls the weighting of compression terms relative to the term that preserves relevant information in the latent space. A similar trade-off parameter is the dimensionality of the latent spaces (in principle, $\mathrm{dim}\,(z_x)$ and $\mathrm{dim}\,(z_y)$ need not be the same, though we set them equal in what follows): a larger embedding dimensionality typically allows the model to preserve more information, and a smaller dimensionality enforces stronger compression.  When the dimensionality, the variational form of the compression, and the estimator for the information in the decoder graph are specified, the DVSIB problem becomes a standard deep-learning training problem. In it, the two encoders $q_X$ and $q_Y$ have independent parameters, and they are optimized jointly, end-to-end, together with the NN that estimates the mutual information $I(Z_X;Z_Y)$ in the latent space~\citep{abdelaleem2025deep}. 

\begin{figure}[t]

\begin{tikzpicture}[
  box/.style={draw, rounded corners=2pt, minimum width=2.2cm, minimum height=0.55cm, font=\small, align=center},
  sbox/.style={draw, rounded corners=2pt, minimum width=0.95cm, minimum height=0.55cm, font=\small, align=center},
  arr/.style={-{Stealth[length=4pt]}, thick},
  darr/.style={-{Stealth[length=4pt]}, thick, dashed},
  font=\small
]

\node[box] (fx) at (0,0) {$X = \{F_t,\ldots,F_{t+n_F-1}\}$};
\node[box, below=0.6cm of fx] (concx) {$[\Phi(F_t),\ldots,\Phi(F_{t+n_F-1})]$};
\node[sbox, below=0.6cm of concx, xshift=-1.3cm] (mux) {$\mu(x)$};
\node[sbox, below=0.6cm of concx, xshift=+1.3cm] (sigx) {$\ell(x)$};
\node[box, below=1.2cm of concx, yshift=-0.55cm] (zx) {$Z_X \sim q(z|x)$};
\draw[arr] (fx) -- node[left, font=\footnotesize]{$\Phi$} (concx);
\draw[arr] (concx) -- node[left, font=\footnotesize]{$\Psi_\mu$} (mux);
\draw[arr] (concx) -- node[right, font=\footnotesize]{$\Psi_{\ell}$} (sigx);
\draw[darr] (mux) -- node[left, font=\footnotesize, pos=0.5, xshift=1.5cm, yshift=0.175cm]{$\epsilon\sim\mathcal{N}(0,I)$} (zx);
\draw[darr] (sigx) -- (zx);

\node[box, right=0.25cm of fx] (fy) {$Y = \{F_{t+n_s},\ldots,F_{t+n_s+n_F-1}\}$};
\node[box, below=0.6cm of fy] (concy) {$[\Phi(F_{t+n_s}),\ldots,\Phi(F_{t+n_s+n_F-1})]$};
\node[sbox, below=0.6cm of concy, xshift=-1.3cm] (muy) {$\mu(y)$};
\node[sbox, below=0.6cm of concy, xshift=+1.3cm] (sigy) {$\ell(y)$};
\node[box, below=1.2cm of concy, yshift=-0.55cm] (zy) {$Z_Y \sim q(z|y)$};
\draw[arr] (fy) -- node[right, font=\footnotesize]{$\Phi$} (concy);
\draw[arr] (concy) -- node[left, font=\footnotesize]{$\Psi_\mu$} (muy);
\draw[arr] (concy) -- node[right, font=\footnotesize]{$\Psi_{\ell}$} (sigy);
\draw[darr] (muy) -- node[left, font=\footnotesize, pos=0.5, xshift=1.5cm, yshift=0.175cm]{$\epsilon\sim\mathcal{N}(0,I)$} (zy);
\draw[darr] (sigy) -- (zy);

\begin{scope}[on background layer]
\node[draw, dashed, rounded corners=4pt, fill=gray!8, inner sep=8pt,
  fit=(fx)(fy)(zx)(zy)(concx)(concy)(mux)(sigx)(muy)(sigy)] (sharedbox) {};
\end{scope}

%
\draw[<->, thin, >=Stealth]
  ([yshift=14pt]fx.north west) -- ([yshift=14pt]fx.north east)
  node[midway, above, font=\scriptsize]{$n_F$};
%
\draw[<->, thin, >=Stealth]
  ([yshift=14pt]fy.north west) -- ([yshift=14pt]fy.north east)
  node[midway, above, font=\scriptsize]{$n_F$};
%
\coordinate (nsL) at ($(fx.north west)!0.3!(fx.north east)$);
\coordinate (nsR) at ($(fy.north west)!0.25!(fy.north east)$);
\draw[<->, thin, >=Stealth]
  ([yshift=25pt]nsL) -- ([yshift=25pt]nsR)
  node[midway, above, font=\scriptsize]{$n_s$};

\node[box, below=1.2cm of zx, xshift=2.5cm] (mi) {InfoNCE estimator $\tilde{I}_\mathrm{NCE}(Z_X;Z_Y)$};
\draw[arr] (zy) -- (mi);
\draw[arr] (zx) to[bend right=15] node[left, font=\footnotesize]{$\delta$-predictor} (mi);

\end{tikzpicture}

\caption{\textbf{DySIB architecture}, where boxes represent variables and intermediate quantities, dashed arrows represent sampling, and arrows represent trainable functions implemented via NNs. A video is segmented into a past window $X$ and a future window $Y$ of $n_F$ frames each, and separated by a stride of $n_s$ frames. Each frame is processed by a shared encoder $\Phi$; the per-frame embeddings are concatenated into a delayed embedding and passed through NNs $\Psi_\mu$ and $\Psi_\ell$ that produce the mean $\mu(\cdot)$ and log-variance $\ell(\cdot)$ of a Gaussian posterior $q(z|\cdot)$, from which $Z_X$ and $Z_Y$ are sampled via the reparameterization trick. The weights in the entire shaded block are shared between past and future. A $\delta$-predictor maps $z_x$ to a Gaussian predictive distribution $r(z_y|z_x) = \mathcal{N}(z_x + \mu_\delta(z_x), \mathrm{diag}(\exp(\ell_{\delta}(z_x))))$, parameterizing latent dynamics as residual increments. This predictive distribution serves as the critic for the InfoNCE mutual information estimator. DySIB minimizes $\mathcal{L}_\mathrm{DySIB} = \tilde{I}^E(X;Z_X) + \tilde{I}^E(Y;Z_Y) - \beta\,\tilde{I}_\mathrm{NCE}(Z_X;Z_Y)$, trading compression against predictive information, the latter entirely in latent space.}
\label{fig:architecture}
\end{figure}

\subsection{DySIB: adapting DVSIB to dynamical systems}
\label{sec:dysib}

To apply DVSIB to dynamical systems, we construct paired variables $X$ and $Y$ from past and future segments of high-dimensional trajectories (video recordings). The DVSIB objective is applied to these variables, with $Z_X$ and $Z_Y$ representing latent descriptions of the past and the future. To incorporate appropriate inductive biases, we impose a few structural constraints on the involved NNs (see Fig.~\ref{fig:architecture}), described in detail below. Two of these are particularly important. First, since the same latent variables are expected to describe the system in the past and in the future (approximate time-translation invariance), we tie the encoders: the distributions $q(z_x|x)$ and $q(z_y|y)$ share parameters and are implemented by the same DNN. Second, to respect the differential structure of physical dynamics, we implement a $\delta$-predictor, a small change in the embedding of the past that makes it maximally informative about the immediate future.

\subsubsection{Delayed embeddings and the shared encoder}
\label{sec:dysib_shared_encoder}

We start with observations of a dynamical system as a sequence of high-dimensional frames (for example, image frames of a video) $\{F_1, F_2, \cdots\}$, with each frame $F_t \in \mathbb{R}^D$, where $D$ is the dimensionality of the observation space. At each time $t$, we construct a pair of variables by taking consecutive segments of this trajectory in the observation space. \PGrevision{We define random variables \(X\) and \(Y\) corresponding to past and future observation windows:}
\begin{equation}    
\begin{aligned}
    X &=\{F_t, F_{t+1}, \cdots, F_{t+n_F-1}\}\;,\\
    Y &=\{F_{t+n_s}, \cdots, F_{t+n_s+n_F-1}\}\;,
\label{eq:past_future_frames}
\end{aligned}
\end{equation}
each consisting of $n_F$ consecutive frames, with the first frame of $Y$ shifted by a \emph{stride} of $n_s$ frames relative to the first frame of $X$. Thus, when $n_s<n_F$, the two windows overlap. 
\PGrevision{Here and throughout, uppercase letters denote random variables, while lowercase letters denote their individual realizations.}
Using multiple frames is necessary because a single frame typically does not determine the full state of the system. For example, quantities such as velocity or acceleration cannot be inferred from a measurement at a single time point. A temporal window provides enough context to resolve these variables and is analogous to a standard delayed embedding of the dynamics~\cite{takens1981detecting}.

The dynamics we want to study are approximately time-translation invariant, so that the latent description of a frame should not depend on when the frame occurs. Thus, each frame $F_t$ is mapped by a {\em shared} encoder $\Phi: \mathbb{R}^D \to \mathbb{R}^{d_F}$ into its own latent representation, where $d_F$ is the per-frame embedding dimension. $\Phi$ is an NN whose specific architecture depends on the problem. For a window $X$, this produces a sequence of vectors $\Phi(F_t), \cdots, \Phi(F_{t+n_F-1})$, which we concatenate to form a representation of the past window in $\mathbb{R}^{n_F d_F}$. The same construction is applied to the future window $Y$.

From this concatenated vector we define a distribution over a latent variable $Z \in \mathbb{R}^{k_z}$. Here $k_z$ is the (bottleneck) dimension of the latent representation. In what follows, we will vary both $k_z$ and $n_F$, and we will show how to use this variation to determine the dimensionality and temporal order of the underlying dynamics from data.
We parameterize this posterior distribution as a diagonal Gaussian, specified by its mean and log-variance $\mu(x), \ell(x): \mathbb{R}^{n_F d_F}\rightarrow  \mathbb{R}^{k_z}$, such that
\begin{equation}
    q(z_x | x) = \mathcal{N}\!\left(\mu(x),\, \mathrm{diag}(\exp(\ell(x)))\right),
\label{eq:encoder}
\end{equation}
and similarly for $q(z_y | y)$ with $\mu(y)$ and $\ell(y)$ (see Fig.~\ref{fig:architecture}). The means and the log-variances are given by NNs $\Psi_\mu$ and $\Psi_\ell$ acting on the $n_F d_F$-dimensional embedding. As with $\Phi$, the specific architecture is problem-specific.

Finally, insisting now on approximate past-future symmetry, we use the same per-frame encoder $\Phi$ and posterior parametrization $\mu, \ell$ for both the past $X$ and the future $Y$. As a result, the past and the future are represented in the same latent coordinates, so that our approach now can discover the dynamics of the compressed representations of the system in the latent space. 

In the main analysis below we use $n_s=1$, so the past and future windows overlap when $n_F>1$. Even though the past and the future now share frames, predicting one from the other is still nontrivial. This is because the two windows are passed through the same shared frame-encoders and through the same low-dimensional bottleneck, so that the overlapping frames, which show up in different positions in the past and the future, are encoded differently. We compare to non-overlapping windows in App.~\ref{app:nf_compare}, and the results remain qualitatively similar.

\begin{figure*}[t]
\centering
\includegraphics[width=\textwidth]{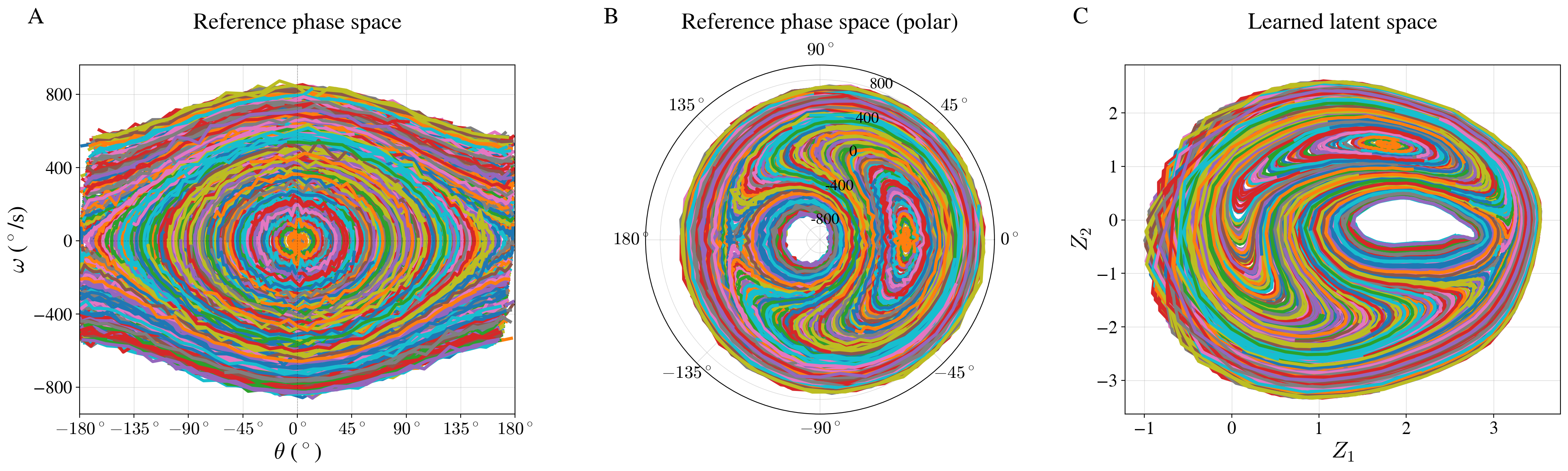}
\caption{\textbf{\PGrevision{Reference} and learned phase space of the pendulum.} \textbf{(A)} Cartesian phase space: angular velocity $\omega$ vs.\ angle $\theta$\PGrevision{, as estimated from the rendered videos (see Sec.~\ref{sec:pendulum})}, for all 1200 trajectories. \textbf{(B)} The same trajectories in the polar projection ($\theta$ as polar angle, $\omega + c$ as radius for a fixed offset $c$), which accounts for the periodic nature of $\theta$ (see Fig.~\ref{fig:si_projection} for the 
\PGrevision{mapping from cylindrical to polar coordinates}).
\textbf{(C)} The two-dimensional latent space learned by DySIB ($k_z = 2$, $n_F = 2$, training videos $N = 1000$, $n_s=1$). The learned latent space closely matches the polar projection, up to a rotation and reflection, recovering the correct phase space structure without supervision.}
\label{fig:phase_recovery}
\end{figure*}

\subsubsection{$\delta$-predictor and DySIB objective}
\label{sec:dysib_delta_predictor}

We are ultimately interested in physical systems whose dynamics can be expressed by ordinary or, in principle, partial differential equations. The laws governing such systems describe rates of change, so that a future state is well approximated by a small differential update to the present~\citep{chen2018neural}. Typical predictive latent-space methods, such as JEPA~\citep{balestriero2025lejepa,maes2026leworldmodel} and information-theoretic dimensionality reduction~\citep{schmitt2023information}, learn a flexible map between an arbitrary past and an arbitrary future, without exploiting this differential structure. However, when the past and future windows are separated by a short time, the future latent $z_y$ should be close to the past latent $z_x$, so it is natural (and likely simpler) to model the increment. We encode this physical inductive bias directly into the architecture and phrase prediction as a construction of a small differential update of the past latent variable. 

Specifically, given a past latent variable $z_x$, we define a conditional distribution over the future latent variable $z_y$, parameterized by an NN we refer to as the $\delta$-predictor. The $\delta$-predictor is an NN, whose specific architecture again is dictated by the structure of the problem. It takes $z_x$ as input and outputs both a mean and a log-variance of the increment,
\begin{equation}
\mu_\delta(z_x), \;\ell_\delta(z_x) \in \mathbb{R}^{k_z}.
\end{equation}
Our predicted future mean is then
\begin{equation}
    z_y^{\text{pred}} = z_x + \mu_\delta(z_x),
\end{equation}
and the overall conditional distribution of the future given the past is
\begin{equation}
r(z_y \mid z_x)
=
\mathcal{N}\!\left(
z_y^{\text{pred}},\;
\mathrm{diag}(\exp(\ell_\delta(z_x)))
\right)\;.
\label{eq:delta_predictor}
\end{equation}
Note that this structure parallels the residual connections used in deep neural networks~\citep{he2016deep}. Thus, in addition to encoding the physical inductive bias of differential dynamics, it is also expected to simplify the training of our relatively complex, inhomogeneous architecture.

This conditional distribution defines the critic for an InfoNCE estimator of the mutual information $I(Z_X; Z_Y)$ between past and future latents, with the residual structure $z_x + \mu_\delta(z_x)$ entering via the critic. This is the quantity we want to maximize. For a pair of observations $(i,j)$, the critic is the log-likelihood of the future latent of the $j$th datum given the past latent of the $i$th one, i.e.,
\begin{multline}
s(z_{x,i}, z_{y,j}) = \log r(z_{y,j} \mid z_{x,i})\\
=
-\frac{1}{2}\sum_{d=1}^{k_z}\left[
\frac{\left(z_{y,j,d} - z_{y,i,d}^{\text{pred}}\right)^2}{\exp(\ell_{\delta,d}(z_{x,i}))}
+ \ell_{\delta,d}(z_{x,i}) + \log 2\pi
\right],
\label{eq:scores}
\end{multline}
where $z_{y,i}^{\text{pred}} = z_{x,i} + \mu_\delta(z_{x,i})$ and $d$ indexes latent dimensions. Overall, the latent predictive information estimator is then
\begin{equation}
\tilde{I}_\mathrm{NCE}(Z_X; Z_Y) =
\frac{1}{B}\sum_{i=1}^{B} \log \left(\dfrac{e^{s(z_{x, i}, z_{y,i})}}{\frac{1}{B} \sum_{j=1}^B e^{s(z_{x,i},z_{y, j}) }}\right),
\label{eq:infonce}
\end{equation}
which compares matched pairs $(z_{x,i}, z_{y,i})$ to mismatched pairs $(z_{x,i}, z_{y,j\ne i})$ within the batch of size $B$.

The full objective for DySIB is then
\begin{equation}
\mathcal{L}_\mathrm{DySIB}
=
\tilde{I}^E(X; Z_X) + \tilde{I}^E(Y; Z_Y)
- \beta\,\tilde{I}_\mathrm{NCE}(Z_X; Z_Y),
\label{eq:dysib}
\end{equation}
where $\tilde{I}_\mathrm{NCE}$ is the InfoNCE instantiation of the general $\tilde{I}^D$ term in Eq.~\eqref{eq:dvsib}, and
\begin{equation}
\tilde{I}^E(X; Z_X)
=
\frac{1}{B}\sum_{i=1}^{B}
D_\mathrm{KL}\!\left(q(z_x \mid x_i)\,\|\,\mathcal{N}(0,I)\right),
\end{equation}
and similarly for $\tilde{I}^E(Y; Z_Y)$. The parameter $\beta$ sets the relative weight of the predictive term and the KL penalty. Throughout, we fix a sufficiently large $\beta=100$ so that the objective is primarily predictive in the latent space, and the dimensionality of the latent variables (which we show later how to choose self-consistently) is the primary regularizer, while the KL penalty terms act only as a weak regularization~\citep{abdelaleem2025deep}. 

The full DySIB architecture, combining the shared encoder, the $\delta$-predictor, and the InfoNCE estimator, is summarized in Fig.~\ref{fig:architecture}.

\section{Reconstructing the phase space of a pendulum}
\label{sec:results}


\subsection{Data and neural architecture}
\label{sec:pendulum}

\PGrevision{We apply DySIB to the pendulum videos from the dataset of \citet{chen2022automated}. These videos were generated by simulating an energy-conserving rigid-rod pendulum and rendering its motion as images, rather than recording a physical experiment. All trajectories correspond to the same pendulum, with identical physical parameters, differing only in their initial conditions and hence their energies.}
The original $128 \times 128$ RGB images are downsampled to $28 \times 28$ grayscale frames, $F_t \in \mathbb{R}^D$ with $D = 784$. Each video consists of 60 frames sampled at $1/60$\,s, covering exactly 1\,s of motion, or about $0.86$ of one small-angle oscillation period ($\approx 1.16$\,s). The dataset comprises 1200 trajectories spanning a range of initial conditions, including both small-angle oscillations and full rotations.

The state of the pendulum, which the videos display, is completely specified by its angle $\theta$ and angular velocity $\omega$, corresponding to a two-dimensional phase space for the dynamical system (Fig.~\ref{fig:phase_recovery}A). The coordinate $\theta$ is periodic with period $2\pi$, giving the phase space the topology of a cylinder, with $\theta$ as the angular coordinate and $\omega$ as the axial coordinate (see Fig.~\ref{fig:si_projection} for an illustration).

Our variational embeddings are expected to map the latent variables into marginal distributions that are approximately standard normal, since the encoder KL term pulls them toward $\mathcal{N}(0, I)$. Thus, to facilitate comparison with the trained representation, we map the cylindrical phase space to a plane by taking $\theta$ as the polar angle and $\omega + c$ as the radius, for a fixed offset $c$ (Fig.~\ref{fig:phase_recovery}B; cf.~Fig.~\ref{fig:si_projection}). In this representation, positive angular velocities map to radii larger than $c$ and negative angular velocities to smaller than $c$, so the two signs of $\omega$ are separated radially. Values of $\omega$ outside the data range leave an empty outer region (large positive $\omega$) and an inner hole (large negative $\omega$). Full rotations with negative $\omega$ trace loops that wind around the hole, while full rotations with positive $\omega$ wind around the hole and the rest of the phase space in the opposite direction. Small oscillations form closed banana-like curves near the stable fixed point.

\PGrevision{The dataset of \citet{chen2022automated} also provides reference values for the angle $\theta$, angular velocity $\omega$, and energy for each frame. These were obtained from the rendered images, rather than directly from the simulation. Specifically, the angle was estimated using OpenCV, and the angular velocity was obtained by fitting a cubic spline to the estimated angles and differentiating it. The supplied labels therefore contain errors introduced by image segmentation and subsequent processing. These labels are not used during training, which is fully self-supervised. To assess the learned representations, we train probe networks on top of the frozen DySIB encoder that map the latent $z$ to these reference physical variables (see App.~\ref{SI:probers}).}
We use up to 1000 trajectories for training and 200 for held-out testing and use two simple probe architectures to test how directly the physical variables are represented in the learned latent space. The first is a random-feature linear probe, which allows only a linear readout after a fixed nonlinear expansion of the latent coordinates. The second is a small MLP probe, which provides a slightly more flexible but still low-capacity readout. These probes are used only for evaluation, not for training DySIB, and their simplicity is intended to check that $\theta$ and $\omega$ are already readily accessible from the latent variables, rather than reconstructed only by a highly expressive supervised model.

For this dataset, we make the following architectural choices. The frame encoder $\Phi$ is a three-layer fully connected multi-layer perceptron (MLP) with hidden width 256 and ReLU activations. The heads $\Psi_\mu$ and $\Psi_\ell$ that produce the mean and log-variance of $q(z|x)$ are taken to be parallel linear maps $W_\mu$ and $W_\ell$; for this simple problem, linear heads already produce nearly perfect results. The $\delta$-predictor is a three-layer fully connected MLP with hidden width 64 and ReLU activations. In more complex scenarios, or for longer temporal context $n_F$, linear heads can become limiting; replacing them with deeper, non-linear $\Psi$ may then be necessary to preserve predictive information and to scale to larger temporal windows. Full implementation details are in App.~\ref{SI:arch}.

For each parameter value, we report the run with the highest test mutual information across ten independent seeds~\citep{abdelaleem2025accurate,gulati2026mutual}. This protocol filters out trials in which stochastic optimization failed to find a good local optimum, so that the reported numbers reflect what the architecture and objective can deliver. All mutual information and probe errors are computed on a held-out test set of 200 trajectories. Probe errors in all panels correspond to this selected run.

\subsection{Self-consistent selection of the latent dimensionality and the temporal window}
\label{sec:model_selection}

The objective in Eq.~\eqref{eq:dysib} depends on the latent dimension $k_z$ and the number of frames $n_F$, which we can determine self-consistently by examining how the estimated mutual information $\tilde{I}_\mathrm{NCE}(Z_X; Z_Y)$ varies with each.

\begin{figure}[htbp]
\centering
\includegraphics[width=\linewidth]{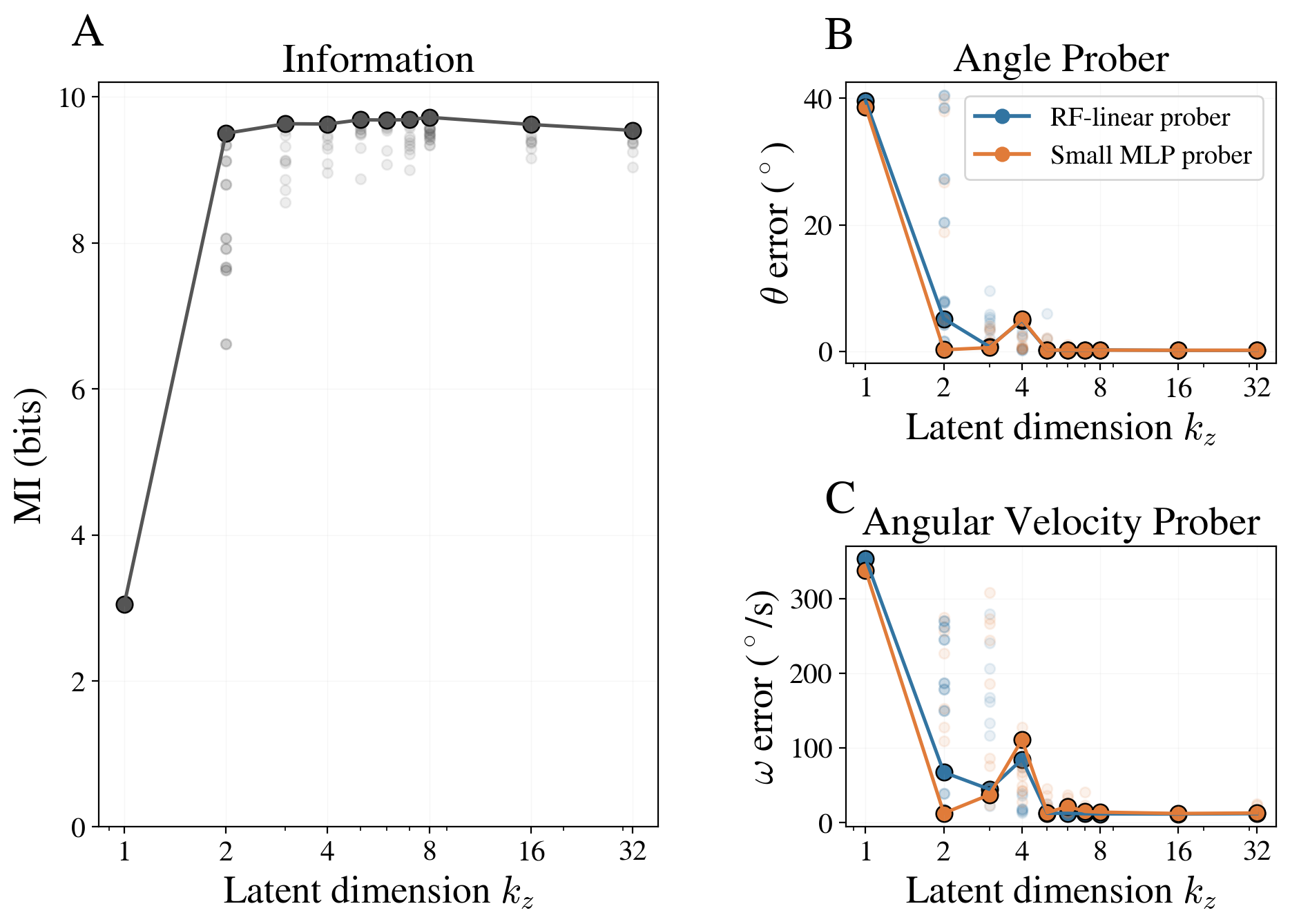}
\caption{\textbf{Mutual information selects the latent dimension.} \textbf{(A)} Mutual information (MI, in bits) as a function of latent dimension $k_z$ with $n_F = 2$ frames, $N = 1000$ training videos, and stride $n_s=1$. MI saturates with no resolvable gain beyond $k_z = 2$, consistent with the two degrees of freedom of the pendulum. In all panels, opaque markers show the run with highest test MI across ten independent seeds; faded markers show the remaining trials. \textbf{(B)} Angle probe RMSE (degrees) vs.\ $k_z$, for a random-feature linear probe (blue) and a two-layer MLP probe (orange). \textbf{(C)} Angular velocity probe RMSE (degrees/s) vs.\ $k_z$, for both probes. Both probe errors drop sharply at $k_z = 2$ and then plateau.}
\label{fig:dim_selection}
\end{figure}

By the data processing inequality, $\tilde{I}_\mathrm{NCE}(Z_X; Z_Y)$ saturates from below at $I(X;Y)$ once $k_z$ is large enough to capture the dynamics, so we take the smallest $k_z$ at which further increments fall within the InfoNCE error bars (computed as in~\citep{abdelaleem2025accurate,gulati2026mutual}) as the latent dimensionality. For the pendulum the saturation at $k_z = 2$ is sharp enough to read off by eye (Fig.~\ref{fig:dim_selection}), consistent with the two-dimensional phase space of the system. This is supported by the probes: both $\theta$ and $\omega$ are accurately recovered at $k_z = 2$, while $k_z = 1$ is insufficient to encode either accurately.

\begin{figure}[htbp]
\centering
\includegraphics[width=\linewidth]{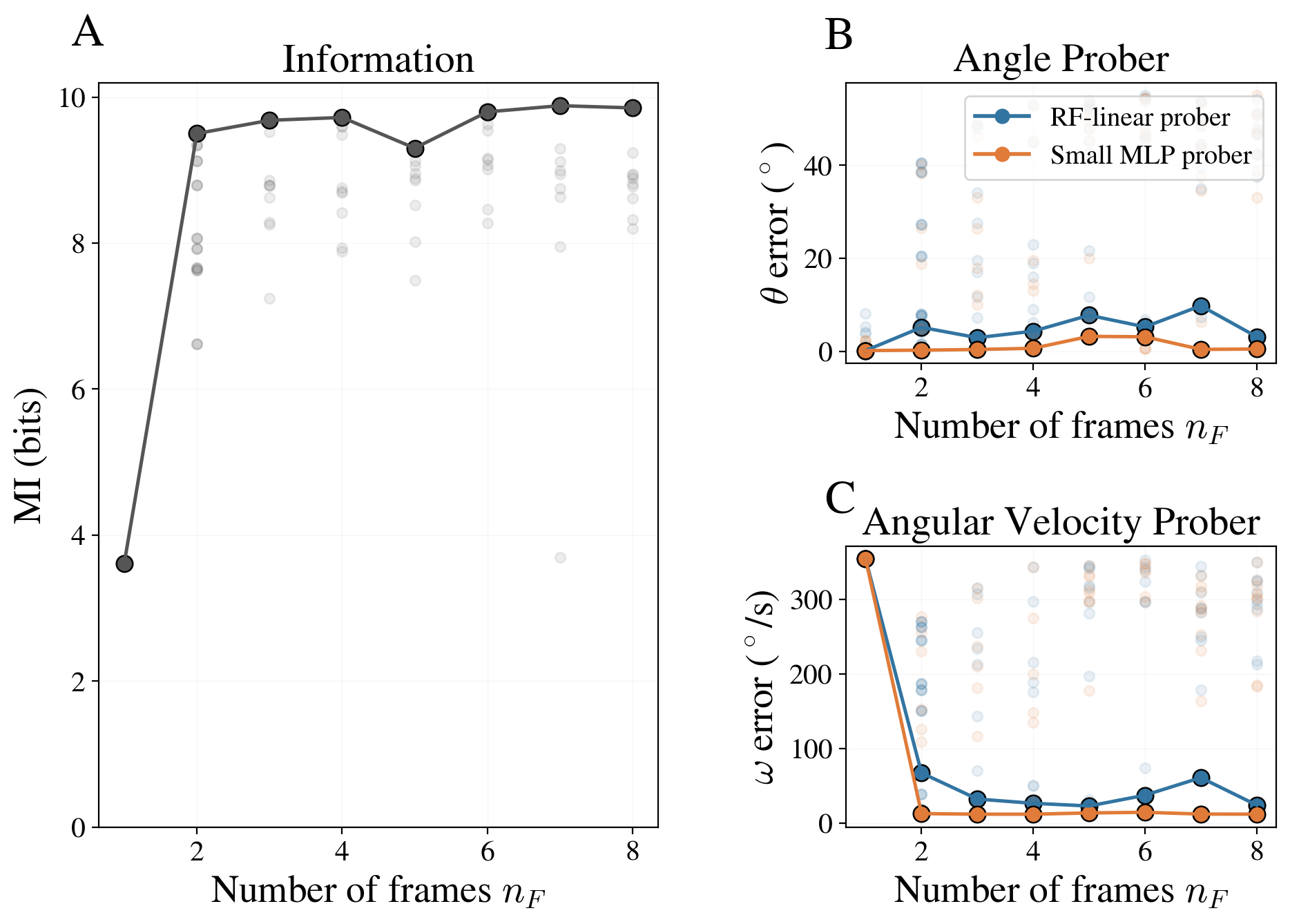}
\caption{\textbf{Mutual information selects the temporal order of the dynamics.} \textbf{(A)} MI (bits) as a function of number of frames $n_F$ with $k_z = 2,$ training videos $N = 1000$, and stride $n_s=1$. MI saturates at $n_F = 2$ consistent with the two frames required to resolve the angular velocity for the pendulum. In all panels, opaque markers show the run with highest test MI across ten independent seeds; faded markers show the remaining trials. \textbf{(B)} Angle probe RMSE vs.\ $n_F$ for a random-feature linear probe (blue) and a two-layer MLP probe (orange). Error is already minimal at $n_F = 1$, as a single frame can accurately recover the angle $\theta$. \textbf{(C)} Angular velocity probe RMSE vs.\ $n_F$, for both probes. Error is large at $n_F = 1$ and drops sharply at $n_F = 2$, since resolving velocity requires at least two frames.}

\label{fig:frame_selection}
\end{figure}

We apply the same procedure to determine the minimal needed temporal window $n_F$ (Fig.~\ref{fig:frame_selection}). The estimated predictive information increases from $n_F=1$ to $n_F=2$ and then shows no resolvable gain for larger windows. This is the expected behavior for a second-order system observed through position-like frames. One frame is enough to infer the angle, but at least two nearby frames are needed to infer the angular velocity. The probes confirm this interpretation. The angle error is already small at $n_F=1$, whereas the angular-velocity error drops sharply only when $n_F=2$. Thus the same saturation criterion that selects $k_z=2$ also selects $n_F=2$, consistent with the temporal order of the pendulum dynamics.

\begin{figure}[htbp]
\centering
\includegraphics[width=\linewidth]{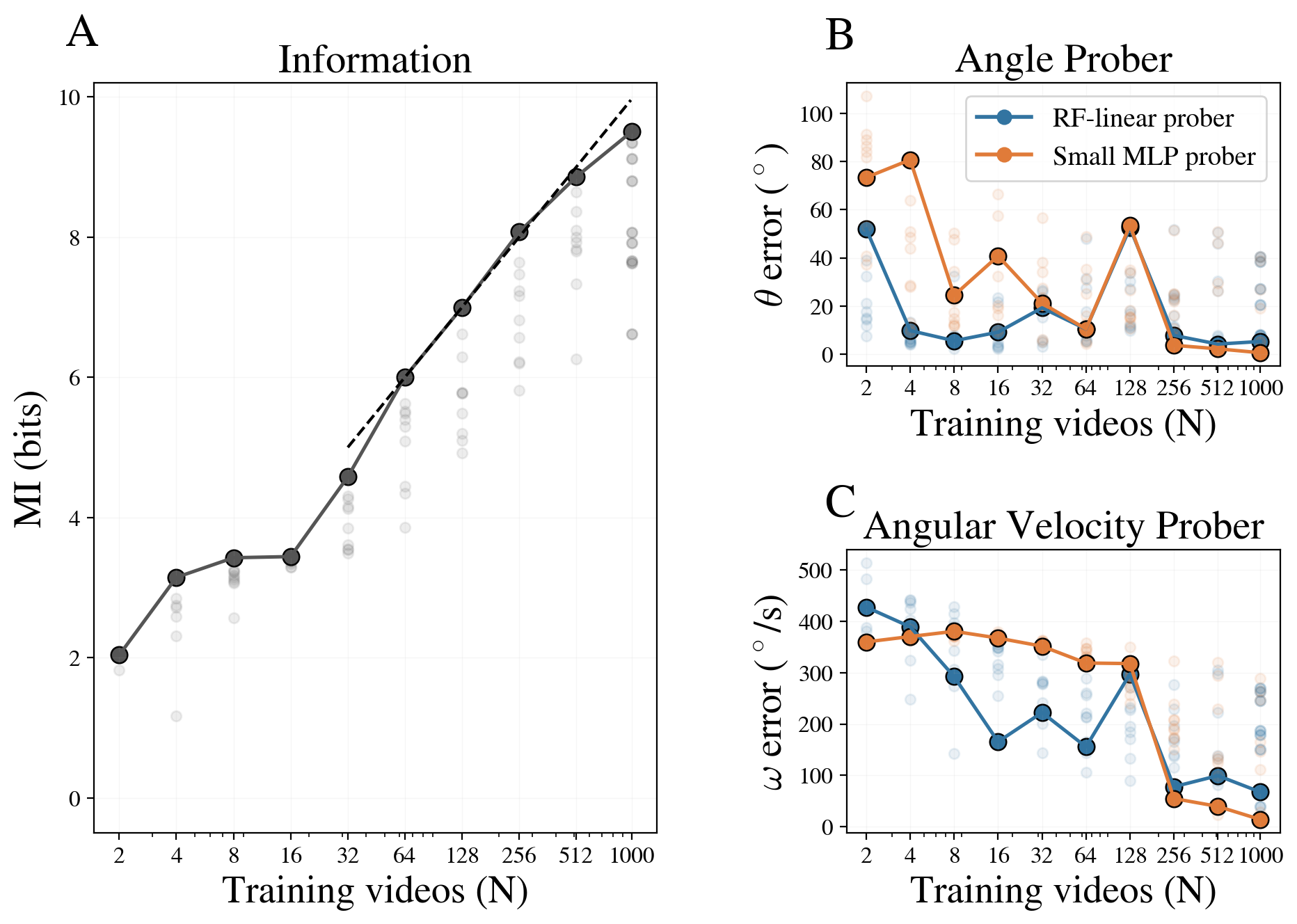}
\caption{\textbf{Sample efficiency of DySIB} ($k_z = 2$, $n_F = 2$, and $n_s=1$). \textbf{(A)} MI (bits) vs. number of training videos $N$ (log scale). MI grows  $\sim\log_2 N$ (dashed line), indicating that independent information scales with the number of distinct initial conditions rather than the total number of frame pairs. As always, opaque markers show the run with highest test MI across ten independent seeds; faded markers show the remaining trials. \textbf{(B)} Angle probe RMSE (degrees) vs.\ $N$, for a random-feature linear probe (blue) and a two-layer MLP probe (orange). Accurate recovery of $\theta$ requires {\em only} $\sim 4$--$8$ videos. \textbf{(C)} Angular velocity probe RMSE (degrees/s) vs.\ $N$. Accurate recovery of $\omega$ requires {\em only} $\sim 256$ videos.}
\label{fig:sample_efficiency}
\end{figure}

We also vary the number of training trajectories $N$ at fixed $k_z = 2$ and $n_F = 2$. The estimated mutual information increases with $N$, while probe errors decrease, with both $\theta$ and $\omega$ recovered accurately using only a fraction of the full dataset (Fig.~\ref{fig:sample_efficiency}). This is consistent with prior work showing that mutual-information-based objectives that operate in the low-dimensional latent representations can capture predictive structure from relatively small datasets~\citep{abdelaleem2025accurate,gulati2026mutual}.

Finally, using a latent dimension larger than necessary does not change the intrinsic dimensionality of the learned representation, but substantially reduces variability across trials (see Fig.~\ref{fig:si_variance} with $k_z=8$), similar to the observation in~\citep{gulati2026mutual}. Intrinsic-dimensionality estimates applied to the $k_z>2$ embeddings remain close to two, showing that the additional latent coordinates do not encode new dynamical degrees of freedom. In practice, we can therefore choose an overparameterized bottleneck, $k_z>2$, and read off the effective dimension from the learned representation. For the pendulum, the representation remains intrinsically two-dimensional, embedded in a larger latent space, and the probes continue to recover $\theta$ and $\omega$ with the same or even improved accuracy across trials.

\subsection{Structure in the latent space}
\label{sec:embeddings}

\begin{figure}[htbp]
\centering
\includegraphics[width=1.05\linewidth]{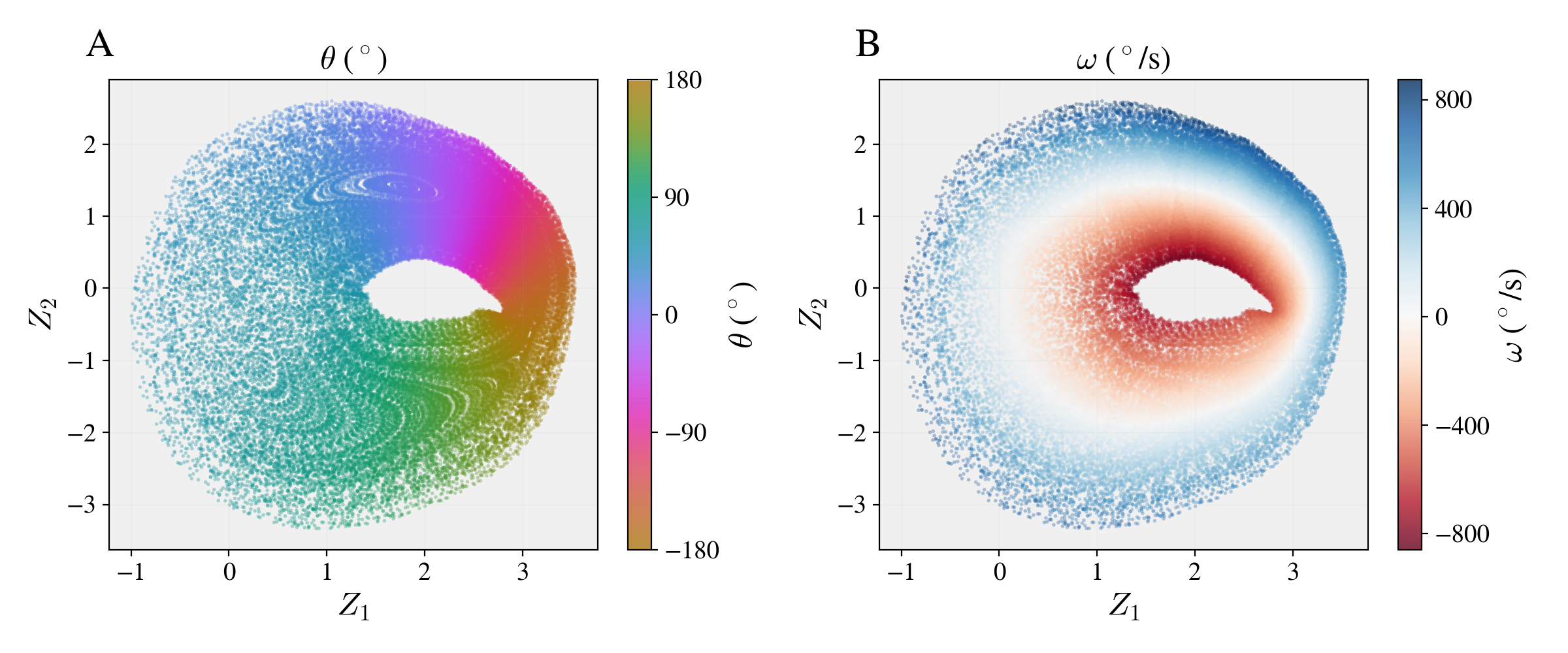}
\caption{\textbf{The learned latent space encodes physical variables.} All trajectories embedded in the learned two-dimensional DySIB latent space for a representative training run ($k_z = 2$, $n_F = 2$, $N = 1000$, $n_s=1$), colored by \textbf{(A)} angle $\theta$ and \textbf{(B)} angular velocity $\omega$, whose values were not used in training (held-out test set). The smooth, globally consistent colorings show that $\theta$ corresponds to the angular coordinate in latent space and $\omega$ to the radial coordinate, consistent with the polar phase space structure. Embeddings colored by kinetic energy, potential energy, and total energy are shown in Fig.~\ref{fig:si_energies}, confirming that the full physical state is encoded in the two-dimensional representation.}
\label{fig:colored_embeddings}
\end{figure}

We examine the structure of the learned two-dimensional latent space by embedding all trajectories (Fig.~\ref{fig:phase_recovery}C) and coloring them by physical variables not used during training (Fig.~\ref{fig:colored_embeddings}).

The angle $\theta$ varies smoothly as the angular coordinate, so that DySIB apparently learned the $2\pi$ periodicity of the problem. The angular velocity $\omega$ varies along the radial direction. Full rotations form closed loops around the origin, while small oscillations form loops around the stable fixed point. This matches the original polar representation of the pendulum phase space (Fig.~\ref{fig:phase_recovery}B) up to global rotations, scaling, and local reparameterizations, consistent with the non-uniqueness of the learned representation up to smooth invertible transformations. The DySIB objective does not enforce any particular shape or orientation of the latent embedding: the predictive information term is invariant under any smooth invertible transformation of the latent variables, and the Gaussian-prior compression 
\PGrevision{constrains the overall}
scale and center. That the learned embedding nonetheless reproduces the textbook polar phase space, including the angular wrap, the radial separation of $\omega$, and the correct locations of the two fixed points, all from raw video alone, is {\em unexpected} and is our {\em central empirical result}.

As expected from the encoding of $\theta$ and $\omega$, this structure extends to other labeled physical quantities (Fig.~\ref{fig:si_energies}). Kinetic energy, which scales as $\omega^2$, follows the same radial organization. Potential energy, which depends on $\cos\theta$, follows the same angular structure. Surfaces of approximately constant total energy align with the trajectories in latent space, consistent with 
\PGrevision{energy conservation in the simulated pendulum dynamics.}

\begin{figure*}[tbp]
\centering
\includegraphics[width=1.75\columnwidth]{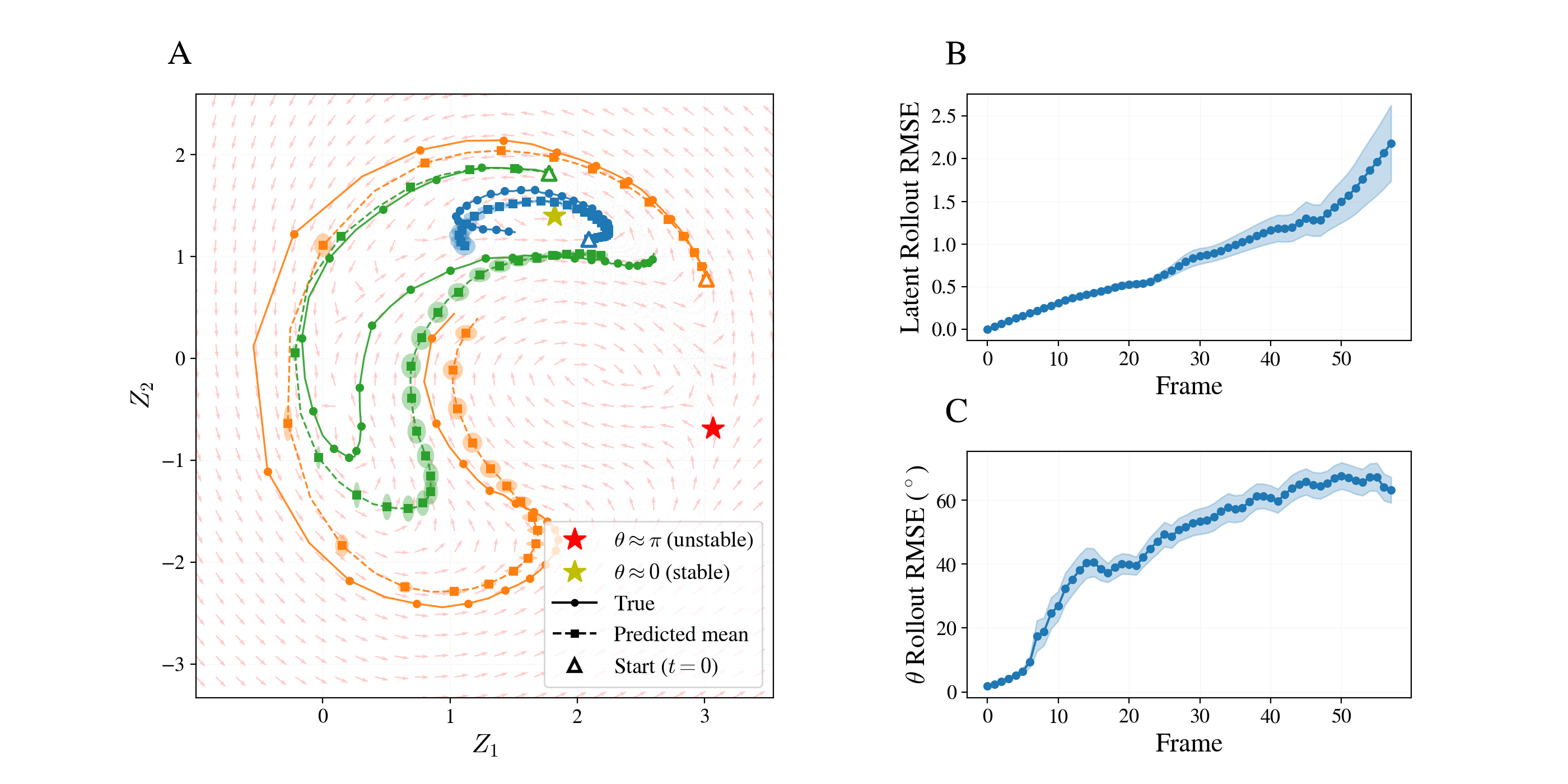}
\caption{\textbf{\PGrevision{Latent dynamics and rollouts}} ($k_z = 2$, $n_F = 2$, $N = 1000$, $n_s=1$). \textbf{(A)} Learned vector field $\mu_\delta(z_x)$ overlaid on the latent phase space, for the same training run as shown in Fig.~\ref{fig:colored_embeddings}. Three representative trajectories are shown: true (solid circles) and predicted mean over stochastic rollout (dashed squares), starting from the triangle marker; shaded regions show $\pm 1$ \PGrevision{SEM} over stochastic samples. Yellow and red stars mark the stable, $(\theta, \omega) \approx (0, 0)$, and unstable, $(\theta, \omega) \approx (\pi, 0)$, fixed points. \textbf{(B)} Latent RMSE as a function of rollout step, averaged over 40 stochastic rollouts; shaded band shows $\pm 1$ \PGrevision{SEM}. \textbf{(C)} Corresponding $\theta$ RMSE in degrees, using the random-feature linear probe. 
\PGrevision{The errors increase with rollout length, showing the limited quantitative forecasting accuracy of the learned update.}
}
\label{fig:forecasting}
\end{figure*}

\subsection{Long-term forecasting}
\label{sec:forecasting}

\PGrevision{Our primary goal is to learn a low-dimensional representation of the dynamics, rather than a model optimized for accurate long-term forecasting. In DySIB, the $\delta$-predictor defines the critic for the InfoNCE objective as it is trained to distinguish matched past-future latent pairs from mismatched ones, rather than to minimize trajectory prediction errors. Nevertheless, the conditional distribution $r(z_y\mid z_x)$ (Eq.~\eqref{eq:delta_predictor}) defines a stochastic update rule in latent space, which we can iterate to examine the dynamics learned by the model. Starting from an initial latent $z(0)$, we generate predicted trajectories $z(t)$ by iterating, for $t=0,1,2,\ldots$:}
\begin{equation}
z(t+1) \sim r(\,\cdot\,|\,z(t)).
\end{equation}

The mean predicted increment $\mu_\delta(z)$ defines a vector field in the latent space (Fig.~\ref{fig:forecasting}A, background quiver plot). The flow has the expected qualitative structure: trajectories orbit around the stable fixed point for oscillatory motion and circulate around the origin for full rotations. The learned flow also shows the expected neighborhoods of the fixed points, corresponding to the stable equilibrium at $(\theta, \omega) = (0, 0)$ and the unstable equilibrium at $(\theta, \omega) = (\pi, 0)$.

\PGrevision{To examine the behavior of this learned update under repeated iteration, we initialize rollouts from the encoded latent states of held-out trajectories}
and generate stochastic trajectories by sampling from $r(z_y \mid z_x)$ at each step. A few representative trajectories and predictions are shown in Fig.~\ref{fig:forecasting}A. We compare these rollouts to the ground-truth evolution over time (Fig.~\ref{fig:forecasting}B,C) for all the trajectories in the test set. The RMS error is computed both in the learned space and in terms of the angular coordinate, using a trained angle probe. 
\PGrevision{The rollout errors grow substantially with prediction horizon, although the representative trajectories in Fig.~\ref{fig:forecasting}A retain the qualitative organization of the underlying dynamics.}
\PGrevision{As implemented, DySIB is a contrastive method for finding embeddings, and its $\delta$-predictor need not be the best forecaster. It is possible, however, that a better predictor may be built on top of the embeddings that DySIB finds. To test this, we trained separate dynamical models on the frozen DySIB latent space. Both a one-step residual MLP and a neural ODE yield smaller rollout errors than the original $\delta$-predictor, with a substantial improvement for the neural ODE (Fig.~\ref{fig:posthoc_roll}; App.~\ref{SI:posthoc_roll}).
}

\section{Discussion}
\label{sec:discussion}
Recovering the dynamical state variables and equations of motion of a system from observations is a long-standing dream and an active research program at the interface of physics and dynamical systems theory~\citep{packard1980geometry,takens1981detecting,eckmann1985ergodic}. In principle, classical methods established that a generic scalar time series, embedded in a delayed-coordinate space of sufficient dimension, can reconstruct the underlying attractor. They also created tools for estimating embedding dimension, Lyapunov spectra, and equations of motion directly from data~\citep{crutchfield1987equations,sugihara1990nonlinear,kennel1992determining}. In practice, however, none of this delivered at a meaningful scale until the advent of big data and large-scale computation. Even then, applications to real experimental observations have remained problem-specific, each requiring a hand-crafted set of candidate variables or a tuned function library to represent the dynamics~\citep{stephens2008dimensionality,schmidt2009distilling,brunton2016discovering,ushio2018fluctuating,daniels2019automated,ahamed2021capturing}. What has been missing is a generic, end-to-end procedure that, given only raw high-dimensional measurements of a system, recovers a low-dimensional, interpretable phase space and the dynamics on it. Here we propose such a procedure,
\PGrevision{demonstrate it on simulated videos of a rigid-rod pendulum, and recover the dimensionality, topology and qualitative dynamical structure of its phase space}
directly from 
\PGrevision{these videos.}

Our work builds on the argument that the dynamics of a system is essentially captured by the mutual information between its past and future, articulated several decades ago~\citep{grassberger1986toward,bialek2001predictability,creutzig2009past}. 
\PGrevision{One challenge has been turning this principle into a practical approach for detecting the latent dynamical variables and for reconstructing the state space and the dynamics on it.}
\PGrevision{DySIB provides one such approach.}
It operationalizes past--future predictive information as a symmetric information bottleneck in which both the past and the future are compressed simultaneously, 
\PGrevision{seeking}
the most compressed summary of the past that retains predictive information about the future. 
In the strict $\beta \to \infty$ limit, 
\PGrevision{if all predictive information is retained, this summary $Z_X$ is a sufficient statistic of the past for the future.}
This recasts the search for effective variables as a data-driven counterpart of the Landau program. Where Landau identifies order parameters from broken symmetries, DySIB 
\PGrevision{seeks to identify effective variables}
from the requirement that they predict their own future. The objective also connects to recent representation-learning work showing the value of latent prediction over reconstruction~\citep{assran2023self,balestriero2025lejepa,maes2026leworldmodel}.

\PGrevision{One potential limitation is that predictive information need not be exclusively due to the dynamical variables, since static or slowly varying features (nuisance) in the observations can also be highly predictive. In subsequent work~\cite{gulati2026contrast}, we study this in controlled noise settings and demonstrate how appropriately designed objectives to approximate conditional mutual information can be used to suppress such static features. How to isolate the information relevant to the dynamics in more general settings remains an open question, though Ref.~\cite{gulati2026contrast} suggests an approach based on the distinguishable time scales of the dynamics and of the nuisance features.}

The same information-theoretic objective also yields a criterion for learning the dimensionality and temporal order of the dynamics. The estimated predictive information saturates with $k_z$ and $n_F$ once the latent space and the temporal window are large enough to capture the system's intrinsic dynamics. For the pendulum, this happens at $k_z = 2$ and $n_F = 2$, the right answers given the two degrees of freedom of the system and the requirement of at least two frames to resolve angular velocity. The larger-$k_z$ experiments further show that overparameterizing the bottleneck need not add spurious dynamical dimensions, and the learned manifold remains intrinsically two-dimensional, with a smaller trial-to-trial variability.

The agreement between the true and learned phase portraits (Fig. 2B,C) is structural, going well beyond visual resemblance. \PGrevision{Both portraits exhibit the same skeleton with a stable equilibrium of oscillatory character, an unstable saddle, separatrices that partition the plane into oscillations and full rotations, and an annular topology in which the two directions of rotation wind around a common inner hole. The} decodability of $\theta$ and $\omega$ from the latent (Sec.~\ref{sec:embeddings}) further suggests that the correspondence is realized by a smooth change of coordinates, although we make no claims that periods or conserved quantities were learned quantitatively correctly. \PGrevision{What we do claim is that DySIB recovers the qualitative organization of the phase space, with the dynamical regimes embedded in their proper topological relation.}

The phase space of a dynamical system is itself defined only up to smooth invertible reparameterizations of its coordinates. The latent space recovered by DySIB inherits this gauge freedom: any smooth invertible transformation of $Z_X$ and $Z_Y$ leaves the predictive mutual information $I(Z_X; Z_Y)$ unchanged. The Gaussian-prior compression terms partially fix the gauge by 
\PGrevision{constraining}
the overall scale and centering of the latent distribution, but they do not single out a unique representation. 
\PGrevision{While the variational posterior is Gaussian for each input, the aggregate posterior is a mixture of Gaussians. The choice of prior favors but does not enforce Gaussianity, allowing for nontrivial structure such as the hole in the learned embedding (Fig.~\ref{fig:phase_recovery}C).}
Even imposing exact marginal Gaussianity leaves a large residual symmetry, the same indeterminacy that makes nonlinear \PGrevision{independent component analysis (ICA)} unidentifiable~\citep{hyvarinen1999nonlinear}. This makes it hard to compare two learned latent spaces directly. A method that learns the right dynamics may present them in coordinates that look very different from another method's, or from those of another random seed of the same method. A canonical form into which any equivalent representation could be transformed would resolve this. One natural candidate, in the spirit of recent geometry-based reconstruction approaches~\citep{yair2017reconstruction}, is to diagonalize the linear part of the latent transition operator and order its eigenvalues, giving a normal-mode coordinate system. Whether this normal-mode form is enough in general, or whether a more flexible construction such as a normalizing flow~\citep{li2020neural} on top of the encoder is needed, remains open. In the present work, the probe networks substitute for a canonical form as the decodability of $\theta$ and $\omega$ from the latent confirms correctness in a representation-independent way. When the right physical coordinates are not known a priori, this falsifiability test is unavailable, and validation must come from out-of-distribution forecasting, internal consistency, or comparison with downstream symbolic regression.

Many of the architectural choices we made in this work are not essential to the underlying principle of {\em selecting variables by their predictive content in the latent space}. The $\delta$-predictor encodes the physical expectation that consecutive latent states differ by a small increment, an inductive bias that clearly helps for the pendulum but that could be relaxed for systems whose dynamics are not naturally written as small differential updates. Similarly, the use of overlapping past and future windows with $n_s=1$ is an implementation choice for estimating a one-frame latent update and not an essential part of the objective. The linear projections from the concatenated frame embeddings to the latent mean and log-variance worked well here only because the underlying dynamics are this simple, and richer observations or longer temporal context would presumably demand deeper nonlinear maps. Our use of the InfoNCE lower bound on the predictive mutual information was likewise 
\PGrevision{an implementation choice. The bound is limited by $\log_2 B$ (see App.~B), while its evaluation scales as $O(B^2 k_z)$, where $B$ is the batch size. If the predictive information grows extensively with the number of dynamical variables, its estimation using InfoNCE would thus require exponentially large batch sizes. Other estimators of mutual information may be more appropriate for such systems.}

Our choice of the simple rigid-rod pendulum as an \PGrevision{initial demonstration} of DySIB was deliberate. \PGrevision{With its reference estimates of angle and angular velocity (subject to errors from image segmentation and subsequent processing)} available for every frame, with deterministic and low-dimensional dynamics, and with clean \PGrevision{simulated} video, the pendulum is \PGrevision{a controlled} setting in which DySIB's success or failure can be \PGrevision{assessed directly}. Our point was to verify that the method recovers the right physics where we already know what right means. Even the video itself is forgiving in this example: $28\times 28$ pixels are sufficient, performance degrades in our experiments only below roughly $16\times 16$ pixels, and resolutions above $28\times 28$ give no further improvement for this problem. \PGrevision{The present results do not establish robustness to experimental observation noise or temporally persistent nuisance features~\citep{gulati2026contrast}, nor do they establish applicability to more complicated dynamics. More demanding tests,} namely systems with strong noise, chaotic dynamics, or multiscale separation, and ones where the dynamical state variables are not known a priori, \PGrevision{remain for future work}. Another next step should be to feed the learned latents into symbolic regression or sparse equation discovery~\citep{schmidt2009distilling,brunton2016discovering,daniels2019automated,gurevich2024spider}, closing the loop from raw observations to interpretable equations of motion expressed in coordinates that the system itself selects. 


\begin{acknowledgments}
This work was funded, in part, by the Simons Foundation Investigator grant to IN and by the National Science Foundation Grant No.~2409416. PG was additionally funded by the Tarbutton Interdisciplinary Postdoctoral Fellowship at the Emory College of Arts and Sciences. \PGrevision{PG and IN were also supported, in part, by the U.S. Department of Energy, Office of Science, Genesis Mission, under Award No.~DE-SC0026711.}  We acknowledge support of our work through the use of the HyPER C3 cluster of Emory University's AI.Humanity Initiative. IN is grateful to Kristofer Bouchard and Vincenzo Vitelli for illuminating discussions. 

\end{acknowledgments}

\section*{Code availability}
The code to set up and train DySIB, as well as the analyzed data from trained runs to reproduce all the figures in this paper is available at \texttt{https://github.com/paarthgulati/DYSIB\_Pendulum}.

\section*{Use of AI tools}
\PGrevision{AI tools were used to assist with editing the manuscript and with writing and managing the code for the training and analysis workflows. The authors reviewed and verified all AI-assisted text and code, and take full responsibility for the content of this article.}

\bibliography{ref}

\appendix
\setcounter{figure}{0}
\renewcommand{\thefigure}{S\arabic{figure}}
\label{sec:SI}

\section{Implementation and evaluation}
\label{SI:implementation}
\subsection{Architecture}
\label{SI:arch}
The shared encoder $\Phi$ is a three-layer MLP with hidden width 256 and ReLU activations that maps each frame $F_t \in \mathbb{R}^{784}$ to a per-frame embedding of dimension $d_F = 32$. The concatenated delayed embedding $[\Phi(F_t), \ldots, \Phi(F_{t+n_F-1})] \in \mathbb{R}^{32 n_F}$ is passed through two parallel linear heads $W_\mu$ and $W_\ell$ producing the mean $\mu(x) \in \mathbb{R}^{k_z}$ and log-variance $\ell(x) \in \mathbb{R}^{k_z}$, from which $z_x$ is sampled via the reparameterization trick. The networks are shared with $z_y$. The $\delta$-predictor is a three-layer MLP with hidden width 64 and ReLU activations mapping $z_x \in \mathbb{R}^{k_z}$ to the mean increment $\mu_\delta(z_x) \in \mathbb{R}^{k_z}$ and log-variance $\ell_\delta(z_x) \in \mathbb{R}^{k_z}$. All linear layers are initialized with Xavier uniform initialization.

\subsection{Training}
\label{SI:training}
We train on the \PGrevision{simulated} pendulum dataset \citep{chen2022automated}, using up to the first 1000 videos for training and the final 200 for held-out evaluation. We downsample the original $128 \times 128$ RGB frames to $28 \times 28$ grayscale ($D = 784$); each video contains $T = 60$ frames, resulting in $T - n_s - n_F + 1$ valid past-future pairs per trajectory. We train with the Adam optimizer with a batch size of 1024 for 300 epochs at a learning rate of $10^{-4}$. The trade-off parameter $\beta$ is set to 100, making the KL penalty small relative to the predictive mutual information term $\beta\,\tilde{I}_\mathrm{NCE}(Z_X; Z_Y)$ and ensuring the objective is primarily driven by predictive information in latent space, aiming for the compressed representations to be nearly sufficient statistics for the dynamics. Then the role of the KL penalty is largely to constrain the latent embeddings to be marginally near-standard Gaussian. We run at least ten independent seeds per configuration and select the run with highest test mutual information. Code is available at \texttt{https://github.com/paarthgulati/DYSIB\_Pendulum}.

\subsection{Evaluation and probe networks}
\label{SI:probers}

We train probe networks with the frozen DySIB encoder to predict 
\PGrevision{reference}
physical quantities, a standard evaluation strategy for latent-space models~\citep{maes2026leworldmodel}. The probes take the posterior mean $\mu(x)$ as input and are trained to predict the physical state at the start of the past window. They test how simply the physical variables can be read out from the learned latent space.

We use two probe architectures. The first is a random-feature linear probe~\citep{rahimi2007random}. The latent vector is passed through a fixed random linear map from $k_z$ to a higher-dimensional space, followed by a Softplus nonlinearity, and only the final linear readout is trained. The random-feature weights are not learned and are drawn independently from a zero-mean Gaussian with standard deviation $2/\sqrt{k_z}$, the variance-preserving initialization for Softplus activations. The linear readout is fit by closed-form ridge regression with regularization $\lambda=10^{-3}$.

The second probe is a small MLP with two hidden layers and ReLU activations, followed by a linear output layer. The MLP probes are trained with DySIB weights frozen for 100 epochs using Adam at a learning rate $10^{-4}$.

Separate probes are trained for $\theta$ and $\omega$. For $\theta$, the probe predicts $(\cos\theta,\sin\theta)$, and the angle is recovered using $\mathrm{atan2}$; errors are computed modulo $2\pi$. For $\omega$, the probe predicts a scalar angular velocity. Training optimizes MSE in the corresponding output space, and performance is evaluated on the held-out 200 trajectories using RMSE, in degrees for $\theta$ and degrees/s for $\omega$.

In the main text, we show results for the random-feature probe with 512 random features, and the MLP probe with 128 nodes in the hidden layers. Results with other probe sizes are in App.~\ref{SI:prober_architecture}.

\subsection{Architectural choices}
\label{SI:arch_choices}
We deliberately chose a simple architecture throughout to isolate the contribution of the objective from architectural improvements. This simplification was sufficient for the pendulum system studied here. Convolutional encoders, transformer architectures suited to complex spatiotemporal data, higher-resolution frames, and nonlinear aggregations of the per-frame embeddings could all improve performance, particularly for more complex or higher-dimensional systems.

\section{Variational bounds on the DySIB objective}
\label{SI:variational}

The DySIB objective, Eq.~\eqref{eq:dysib}, trades off two competing goals: keeping the latent representation $Z_X$ as simple as possible while making it maximally predictive of the future latent $Z_Y$. This is an instance of the information bottleneck \citep{tishby2000information,friedman2001multivariate,abdelaleem2025deep}: compress the input as much as possible while retaining only the information relevant for a target variable. We seek variational approximations for two kinds of terms: the encoder terms $I^E(X; Z_X), I^E(Y; Z_Y)$, which we minimize to ensure compression, and the predictive term $I_\mathrm{NCE}(Z_X; Z_Y)$, which we maximize to ensure that the past is informative about the future. Detailed derivations of all bounds used below are given in~\citep{abdelaleem2025deep}; we recap them here only briefly.

\subsection{Encoder terms}
\label{SI:kl}

The encoder terms penalize the complexity of the latent representation by bounding $I(X; Z_X)$ from above. Intuitively, $I(X; Z_X)$ measures how much information the latent code $z_x$ retains about the input $x$. Minimizing it on its own would encourage the encoder to discard everything; the $\tilde{I}_\mathrm{NCE}(Z_X;Z_Y)$ term counteracts this by rewarding representations that retain information predictive of the future. We start with the definition of mutual information:
\begin{equation} 
    I(X; Z_X) = \mathbb{E}_{p(x, z_x)}\left[\log \frac{q(z_x|x)}{p(z_x)}\right], 
\end{equation}
where $p(z_x) = \int q(z_x|x)p(x)\,dx$ is the aggregate posterior. This quantity is intractable because $p(z_x)$ has no closed form. We introduce a variational approximation $\tilde{p}(z_x)$ to $p(z_x)$ and use the non-negativity of the KL divergence, $D_\mathrm{KL}(p(z_x)\|\tilde{p}(z_x)) \geq 0$, to obtain:
\begin{equation}
    I(X; Z_X) \leq \mathbb{E}_{p(x)}\left[D_\mathrm{KL}(q(z_x|x)\,\|\,\tilde{p}(z_x))\right] \equiv \tilde{I}^E(X; Z_X).
\end{equation}

Now, approximating $p(x)$ with its empirical distribution over a minibatch of $B$ samples and choosing $\tilde{p}(z_x) = \mathcal{N}(0, I)$ as the variational prior gives:
\begin{equation} 
    \tilde{I}^E(X; Z_X) = \frac{1}{B}\sum_{i=1}^{B} D_\mathrm{KL}\!\left(q(z_x|x_i)\,\|\,\mathcal{N}(0,I)\right),
\end{equation}
and identically for $\tilde{I}^E(Y; Z_Y)$. This is the standard KL regularization term from the VAE objective \citep{kingma2013auto}, repurposed here as a complexity penalty rather than a reconstruction regularizer. The choice of $\mathcal{N}(0,I)$ as the prior is standard and ensures the bound is tractable; detailed derivation can be found in \cite{kingma2013auto,alemi2017deep,abdelaleem2025deep}.

\subsection{Predictive term}
\label{SI:infonce}

The predictive term $\tilde{I}_\mathrm{NCE}(Z_X; Z_Y)$ lower-bounds the mutual information between past and future latent codes, which DySIB maximizes to encourage the latent representation to capture predictive structure. We start from the conditional factorization of mutual information:
\begin{equation} 
    I(Z_X; Z_Y) = \mathbb{E}_{p(z_x)}\left[D_\mathrm{KL}(p(z_y|z_x)\,\|\,p(z_y))\right], 
\end{equation}
and apply the Donsker--Varadhan (DV) representation \citep{donsker1983asymptotic}, which provides a lower bound on any KL divergence:
\begin{equation} 
    D_\mathrm{KL}(P\|Q) \geq \sup_T\left[\mathbb{E}_P[T] - \log \mathbb{E}_Q[e^T]\right], 
\end{equation}
where the supremum is over all measurable functions $T$ (the Donsker--Varadhan test function, unrelated to the frame count $T = 60$ of App.~\ref{SI:training}), and equality holds when $T^* = \log\frac{dP}{dQ} + c$. Applying this to $D_\mathrm{KL}(p(z_y|z_x)\|p(z_y))$ and integrating over $p(z_x)$ gives a lower bound on $I(Z_X;Z_Y)$ with a learned critic $T(z_x, z_y)$. In practice, the expectation over $p(z_y)$ is approximated using contrastive sampling: for a batch $\{(z_{x,i}, z_{y,i})\}_{i=1}^B$, matched pairs $(z_{x,i}, z_{y,i})$ serve as positive samples and mismatched pairs $(z_{x,i}, z_{y,j\neq i})$ serve as negatives, giving the InfoNCE estimator \citep{oord2018representation, abdelaleem2025accurate}:
\begin{equation} 
    \tilde{I}_\mathrm{NCE}(Z_X; Z_Y) = \frac{1}{B}\sum_{i=1}^{B}\log\frac{e^{T(z_{x,i},\, z_{y,i})}}{\frac{1}{B}\sum_{j=1}^{B} e^{T(z_{x,i},\, z_{y,j})}},
\end{equation}
where $T(z_{x,i}, z_{y,j}) = s(z_{x,i}, z_{y,j})$ is the log-likelihood critic constructed from the $\delta$-predictor as in Eq.~\eqref{eq:scores}. This estimator is a biased lower bound on $I(Z_X; Z_Y)$ that is bounded above by $\log B$~\citep{oord2018representation,poole2019variational}. With a batch size of $B = 1024$ the ceiling is approximately 10 bits, which is sufficient for the $k_z, n_F$ selection procedure used here. This is the estimator used in Eq.~\eqref{eq:infonce} of Sec.~\ref{sec:dysib}. In other applications, larger batch sizes or estimators without a trivial upper bound may need to be used. 

\section{Supplementary results}
\label{SI:additiona_figures_and_results}

\PGrevision{\subsection{The pendulum phase space in cylindrical and polar coordinates}}
\label{SI:phase_space_geometry}

\begin{figure}[htbp]
\centering
\includegraphics[width=\linewidth]{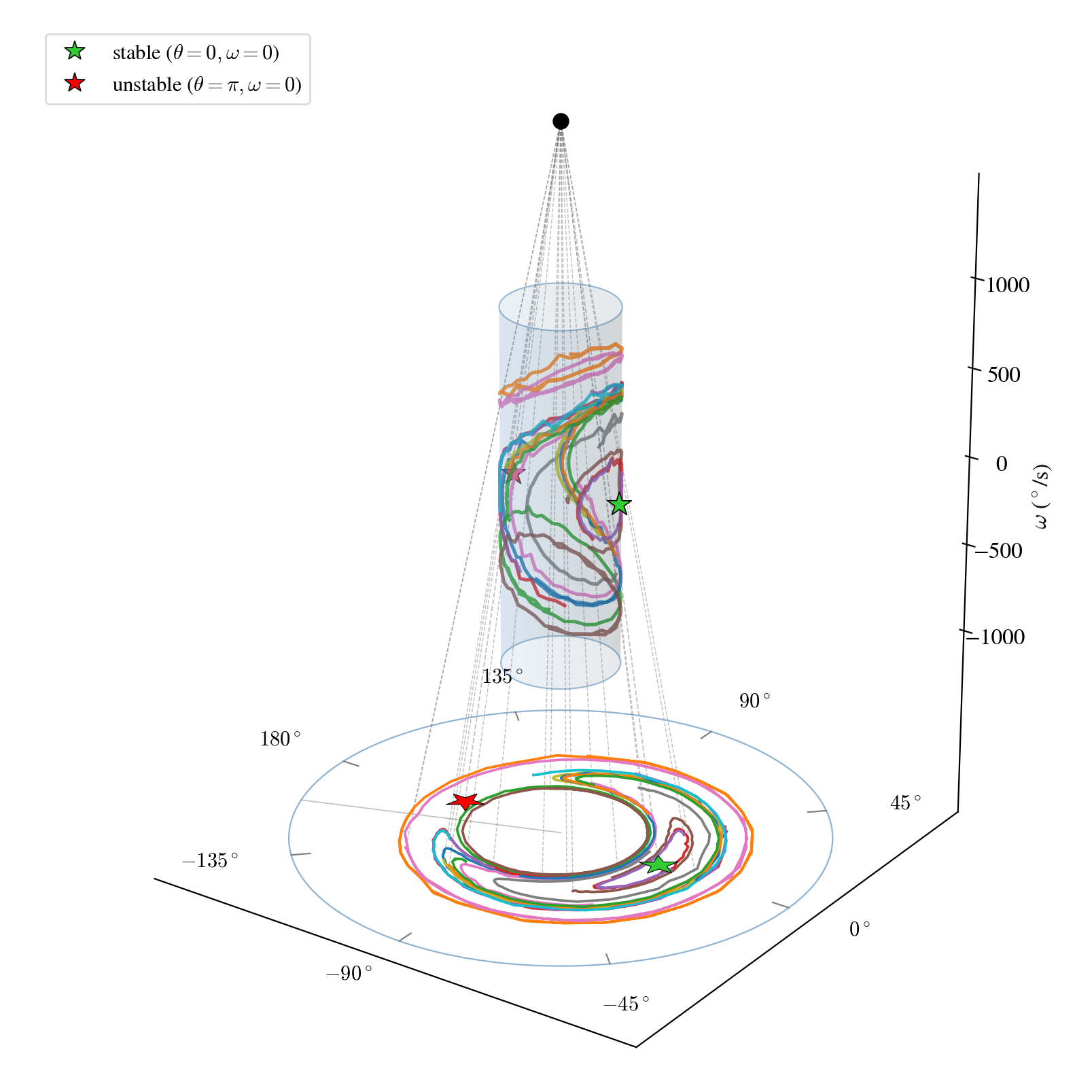}
\caption{\textbf{\PGrevision{Mapping between} the cylindrical and polar phase space representations of the pendulum.} The trajectories wind around the surface of a cylinder, with $\theta$ as the angular coordinate and $\omega$ as the axial coordinate. Projecting onto a polar plane, with $\theta$ as the polar angle and $\omega +c$ as the radius, produces the polar representation recovered by DySIB. The stable fixed point ($\theta = 0$, $\omega = 0$, green star) and unstable fixed point ($\theta = \pi$, $\omega = 0$, red star) are marked on the cylinder. Full rotations appear as closed loops winding around the cylinder and around the origin in the polar projection, while small oscillations form closed curves near the stable fixed point.}
\label{fig:si_projection}
\end{figure}

Because the KL terms in the loss function bias the embeddings to be marginally near-standard Gaussian, the learned latent space recovered by DySIB takes the form of the polar projection of the pendulum phase space, with $\theta$ as the polar angle and $\omega$ as the (offset) radius. Figure~\ref{fig:si_projection} illustrates the 
\PGrevision{mapping between these representations,}
explaining why the learned latent space has the structure it does, including the hole at the origin, the distinction between oscillating and rotating trajectories, and the positions of the two fixed points.

\subsection{Effect of overlap of temporal windows}
\label{app:nf_compare}

\begin{figure}[htbp]
\centering
\includegraphics[width=\linewidth]{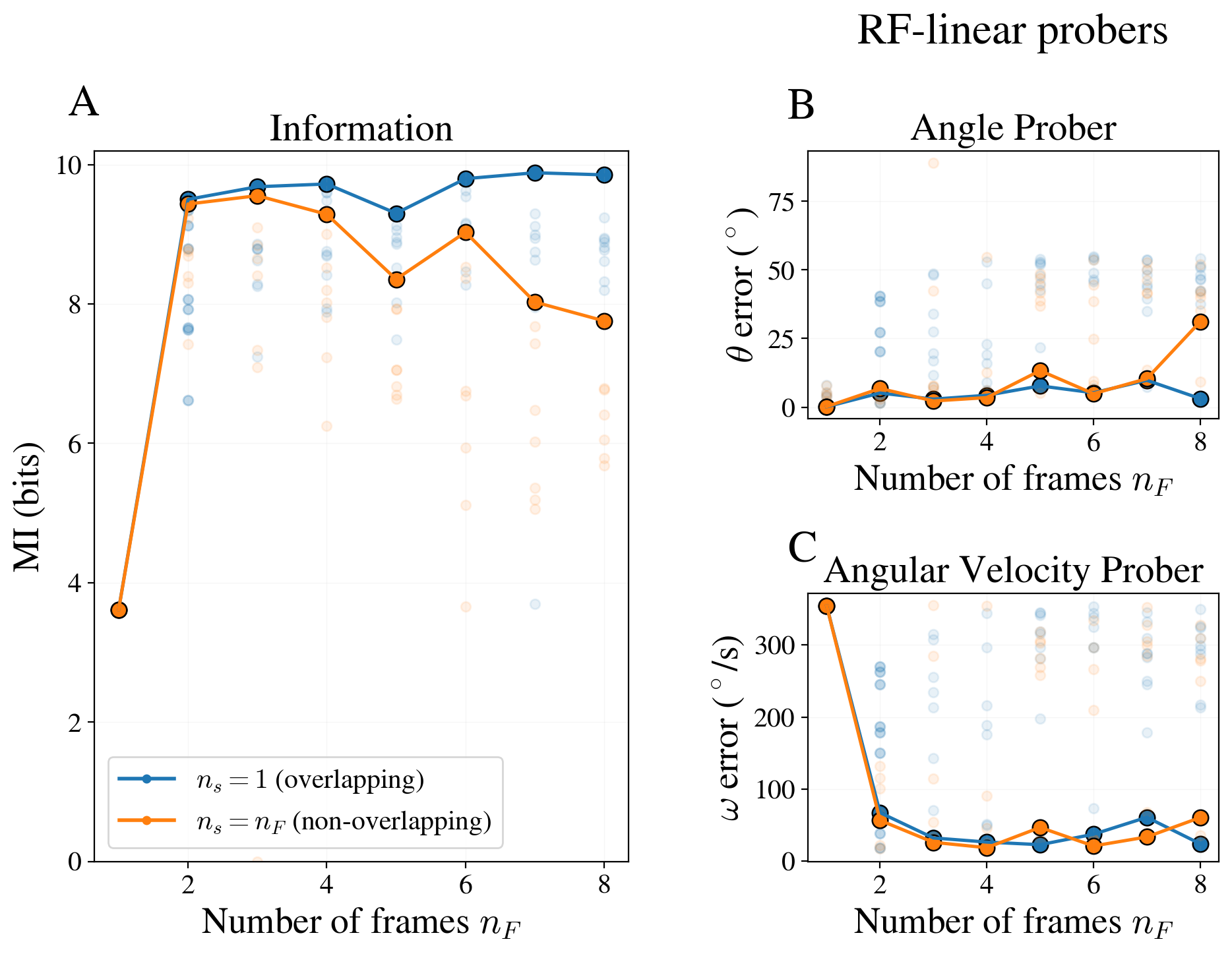}
\caption{
\textbf{Effect of the stride $n_s$.} We compare the overlapping setup used in the main text Fig.~\ref{fig:frame_selection} with non-overlapping past--future windows ($n_s=n_F$) for the sweep against the window length. \textbf{(A)} Estimated predictive mutual information as a function of $n_F$. For $n_s=1$, the information saturates at $n_F=2$, while for $n_s=n_F$ it peaks at $n_F=2$ and slowly decreases for larger windows. \textbf{(B)} Angle probe RMSE and \textbf{(C)} angular-velocity probe RMSE, both computed using the random-feature linear probe. For both choices of $n_s$, the angle error is small throughout, while the angular-velocity error is large at $n_F=1$ and drops sharply for $n_F\geq 2$. Thus the selection of $n_F=2$ is robust to whether the past and future windows overlap.}
\label{fig:si_stride_nf}
\end{figure}

In the main text, we use $n_s=1$, so the past and the future windows overlap when $n_F>1$. This choice corresponds to learning a one-frame update of the delay-coordinate state. To test whether the model-selection result depends on this overlap, we repeat the temporal-window sweep using non-overlapping windows, $n_s=n_F$, and compare it to the main $n_s=1$ setup (Fig.~\ref{fig:si_stride_nf}).

For the overlapping windows used in the main text, the estimated predictive information increases from $n_F=1$ to $n_F=2$ and then saturates, since two frames already suffice to resolve the angular velocity and additional frames add no new dynamical information. For non-overlapping windows, the information instead peaks at $n_F=2$ and then slowly decreases for larger $n_F$, likely because the increasing stride ($n_s=n_F$) requires prediction over longer times (making the $\delta$-predictor less efficient) and reduces the number of past--future pairs per trajectory. This decrease is not reflected in the probe errors. For both choices of $n_s$, the angle error is small throughout, while the angular-velocity error is large at $n_F=1$ and drops sharply once $n_F=2$. Thus both the overlapping and non-overlapping constructions identify two frames as the temporal window needed to resolve the pendulum state. The main conclusion that $n_F=2$ is selected by the dynamics is therefore not an artifact of sharing frames between the past and future windows.

\begin{figure*}[t]
\centering
\includegraphics[width=\textwidth]{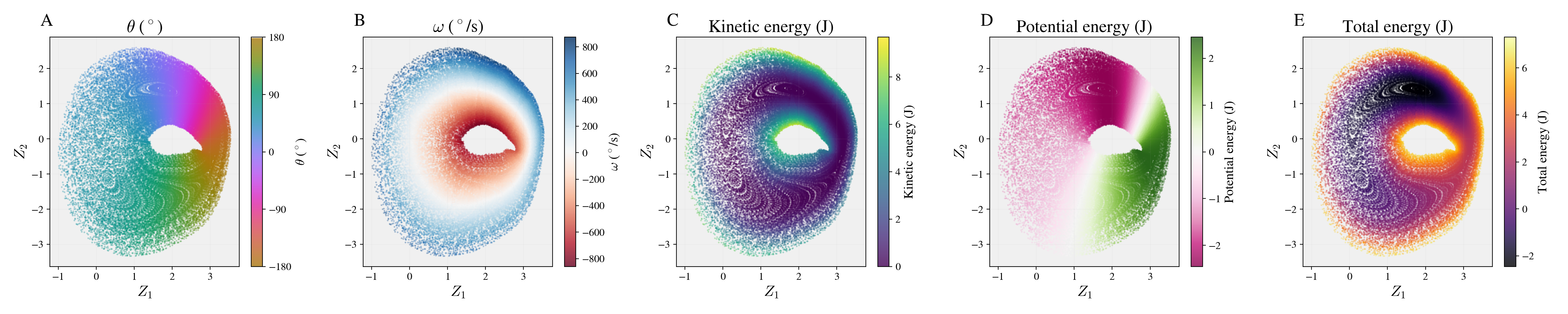}
\caption{\textbf{All tracked physical quantities in the learned latent space} ($k_z = 2$, $n_F = 2$, $N = 1000$). From left to right: angle $\theta$ (degrees), angular velocity $\omega$ (degrees/s), kinetic energy (J), potential energy (J), and total energy (J). Each quantity varies smoothly and consistently with the latent coordinates. Kinetic energy follows the same radial structure as $\omega$ (scaling as $\omega^2$); potential energy follows the same angular structure as $\theta$ (scaling as $1 - \cos\theta$); surfaces of constant total energy align with the trajectories visible in Fig.~\ref{fig:phase_recovery}C, consistent with energy conservation.}
\label{fig:si_energies}
\end{figure*}

\subsection{Other physical quantities in the learned latent space}
\label{SI:all_physical_quantities}
Figure~\ref{fig:colored_embeddings} shows the latent space colored by $\theta$ and $\omega$. Figure~\ref{fig:si_energies} extends this to all tracked physical quantities, demonstrating that the physical state of the pendulum is encoded in the two-dimensional representation. The smooth, globally consistent structure of each coloring shows an embedding closely aligned with the physical state space, not just a useful downstream representation. Together these colorings confirm that the two-dimensional latent space captures the complete physical state of the system.

\subsection{A larger-than-necessary bottleneck reduces trial-to-trial variance}
\label{SI:overparameterization}
In Figs.~\ref{fig:dim_selection},~\ref{fig:frame_selection} the trial-to-trial variance is visible in the $k_z$ and $n_F$ sweeps, particularly at small bottleneck sizes. Figure~\ref{fig:si_variance} shows that training with a larger-than-necessary bottleneck ($k_z = 8$) substantially reduces this variance without changing the saturation point or the intrinsic dimensionality of the learned representation. Nonlinear intrinsic dimensionality estimators applied to the $k_z = 8$ latent space return an effective dimension of approximately 2, confirming that the additional capacity is not used to encode spurious structure. Interestingly, the participation ratio returns a slightly larger value, reflecting the curvature of the learned manifold rather than additional degrees of freedom. This is consistent with the earlier results \citep{gulati2026mutual} that models with a larger-than-necessary bottleneck concentrate the cross-view signal onto an effectively low-dimensional subspace. Note that although the first two principal components of the $k_z = 8$ latent space account for approximately 90\% of the variance (Fig.~\ref{fig:si_variance}E), \PGrevision{the two-dimensional PCA projection does not reproduce the expected annular phase-space structure}
since the angle is encoded nonlinearly. The nonlinear estimators are therefore the appropriate tool for assessing intrinsic dimensionality of the learned latent space. In practice, training with a larger-than-necessary bottleneck and reading off the effective dimension is a robust alternative to sweeping $k_z$.

\PGrevision{
\subsection{Post-hoc forecasting in the learned latent space}
\label{SI:posthoc_roll}}

\PGrevision{
To examine whether the learned representation can support more accurate forecasting, we trained two additional dynamical models using a DySIB encoder trained with $k_z=2$, $n_F=2$, $n_s=1$, and $N=1000$, with its weights fixed at the values learned in the original end-to-end training. Both models were trained on the posterior-mean latent coordinates, without using any physical labels. The first is a one-step residual MLP with two hidden layers of width 64 and $\tanh$ activations, trained to minimize one-step prediction error in latent space. The second is a two-dimensional neural ODE~\citep{chen2018neural}, whose vector field is parameterized by an MLP with the same architecture. Unlike the one-step MLP, the neural ODE was trained using multistep trajectory prediction errors, and its dynamics were integrated using fourth-order Runge--Kutta with two substeps per frame. The models were selected using a validation split within the training dataset.
}

\begin{figure}[htbp]
    \centering
    \includegraphics[width=\linewidth]{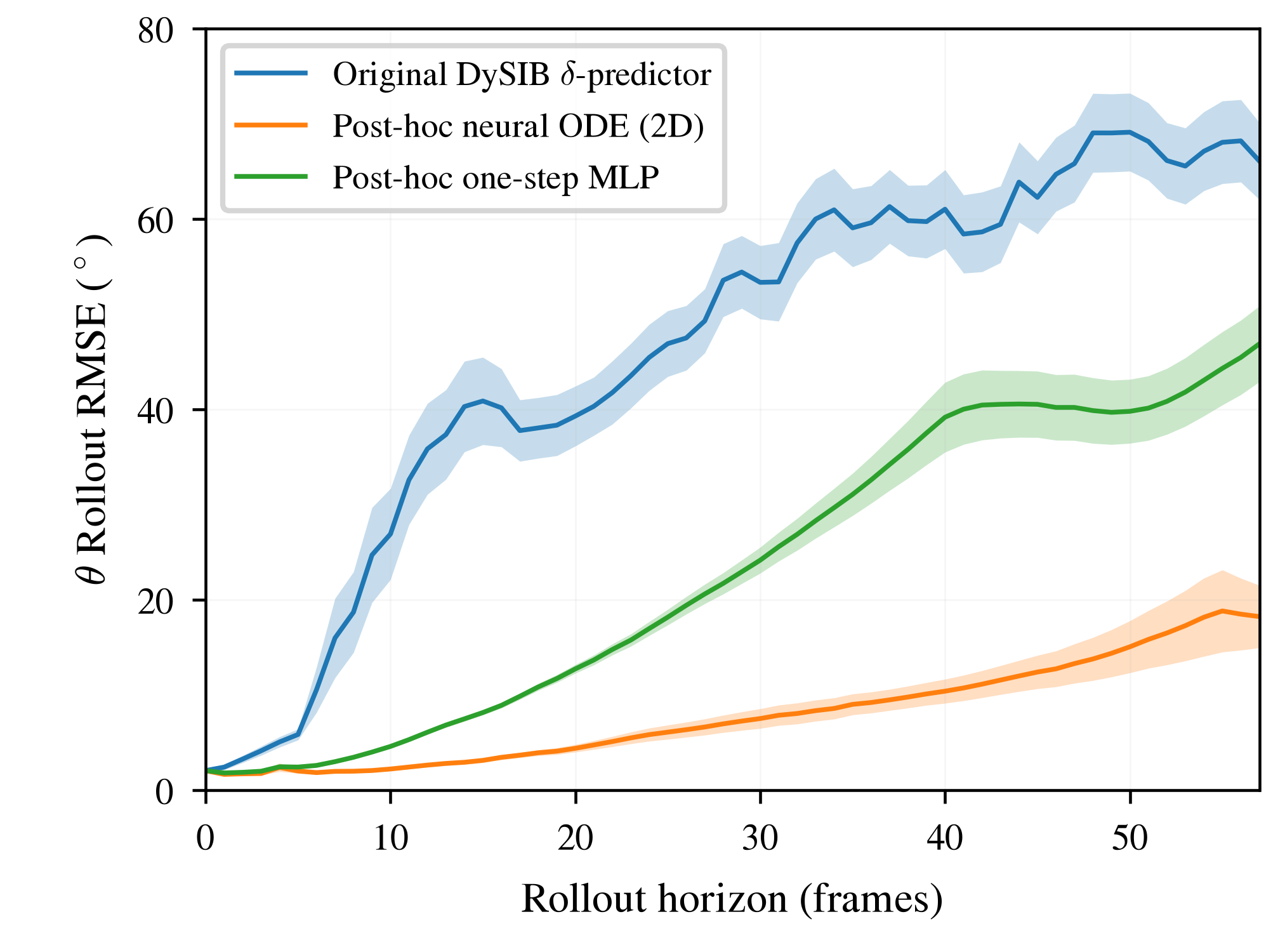}\caption{\PGrevision{\textbf{Post-hoc forecasting in the frozen DySIB latent space.} Angular rollout RMSE over 200 held-out pendulum trajectories for the original jointly trained DySIB $\delta$-predictor, a post-hoc two-dimensional neural ODE, and a post-hoc residual MLP trained on one-step latent prediction error. All methods use the same frozen encoder, initial latent states, and angle probe. The original predictor is evaluated by averaging 40 stochastic latent rollouts before decoding, while the post-hoc predictors are deterministic. Shaded bands indicate $\pm 1$ SEM across trajectories.}}
    \label{fig:posthoc_roll}
\end{figure}

\PGrevision{
Figure~\ref{fig:posthoc_roll} compares these models with the original jointly trained DySIB $\delta$-predictor on the same 200 held-out trajectories. All three methods start from the same encoded initial states, and their predictions are converted to physical angles using the same frozen probe. Both post-hoc predictors improve the rollout accuracy, with the neural ODE producing substantially smaller errors. The remaining errors may reflect inaccuracies in the initial latent representation or the fitted dynamics, particularly for trajectories approaching the unstable fixed point and separatrix, where small errors can lead to qualitatively different trajectories. These results show that the learned representation supports considerably more accurate forecasting than is obtained by directly iterating the contrastively trained $\delta$-predictor.
}

\subsection{Evolution of the learned embedding with training set size}
\label{SI:embedding_vs_N}

\begin{figure*}[t]
\centering
\includegraphics[width=0.8\textwidth]{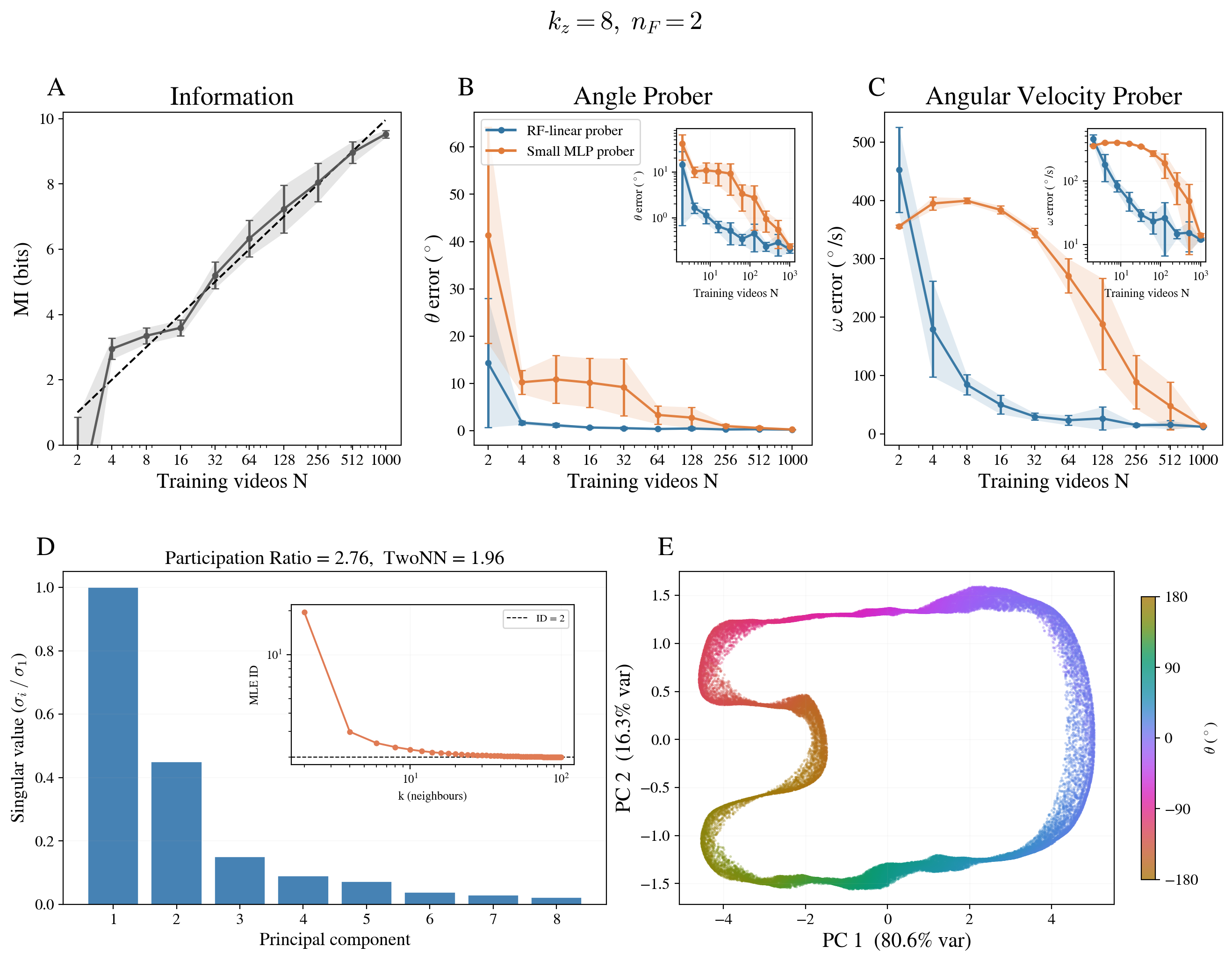}
\caption{\textbf{A larger-than-necessary bottleneck reduces trial-to-trial variance.} 
\textbf{(A--C)} Past-future latent mutual information (bits), angle probe RMSE (degrees), and angular velocity probe RMSE (degrees/s) as functions of training set size $N$, for $k_z = 8$, $n_F = 2$, and $n_s=1$. Error bars show standard deviation across ten independent seeds. Compared to the $k_z = 2$ results (Figs.~\ref{fig:dim_selection},~\ref{fig:frame_selection}), trial-to-trial variability is substantially reduced across all values of $N$. The dashed line in (A) corresponds to $\log_2 N$. In panels (B, C), errors are shown for both the random-feature and the two-hidden-layer MLP probes used in the main text; both give comparable results for the overparameterized bottleneck.
\textbf{(D)} Singular value spectrum of the $k_z = 8$ latent space (representative trial with $N=1000$), with intrinsic dimensionality estimated by three independent methods: Levina--Bickel MLE \citep{levina2004maximum}, TwoNN \citep{facco2017estimating}, and participation ratio (PR) from the cross-covariance singular value spectrum \citep{gulati2026mutual}. The nonlinear estimators (TwoNN, Levina--Bickel) return $d_\mathrm{eff} \approx 2$, while the PR returns a slightly larger value, reflecting the curvature of the learned manifold. \textbf{(E)} The first two principal components of the $k_z = 8$ latent space, colored by $\theta$, show that the angle is encoded in the representation, though not in a linearly separable way.}
\label{fig:si_variance}

\vspace{2em}

\includegraphics[width=0.8\textwidth]{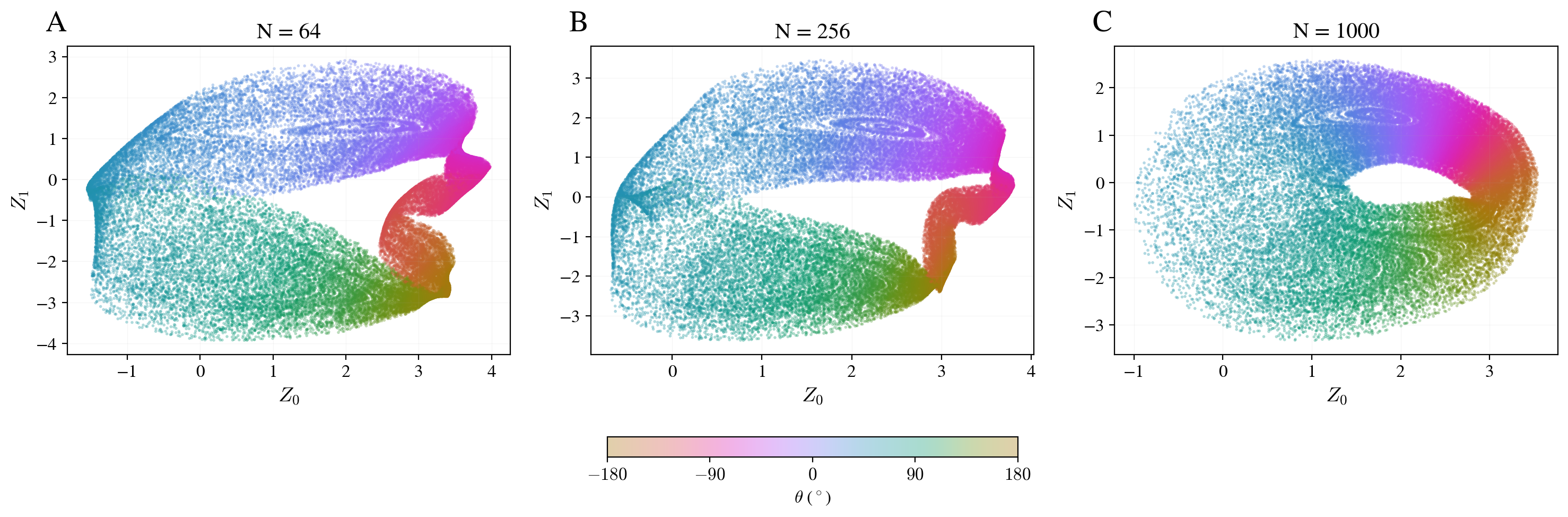}
\caption{\textbf{Evolution of the learned latent space with training set size} ($k_z = 2$, $n_F = 2$, $n_s=1$). Each panel shows all available trajectories embedded in the learned two-dimensional latent space, colored by \PGrevision{reference} angle $\theta$, for $N = 64$ (left), $N = 256$ (center), and $N = 1000$ (right). With $N = 64$, the structure is distorted and the coloring is inconsistent. With $N = 256$, the polar structure begins to emerge. With $N = 1000$, the embedding closely matches the polar phase space of Fig.~\ref{fig:phase_recovery}B, with smooth and globally consistent coloring. The best run across ten seeds is shown for each $N$.}
\label{fig:si_embedding_vs_N}
\end{figure*}

Figure~\ref{fig:si_embedding_vs_N} shows how the qualitative structure of the learned latent space changes as the number of training videos $N$ increases. With very few videos the structure is distorted and the coloring by $\theta$ is inconsistent. As $N$ increases, the embedding progressively organizes into the correct polar structure, with smooth and globally consistent coloring by $\theta$ emerging around $N \sim 256$ and converging by $N = 1000$. This qualitative progression mirrors the quantitative probe error decay in Fig.~\ref{fig:sample_efficiency}, illustrating 
\PGrevision{how the learned phase-space structure emerges as the training set grows.}

\subsection{Effect of probe architecture}
\label{SI:prober_architecture}

\begin{figure*}[tbp]
    \centering
    \includegraphics[width=0.8\linewidth]{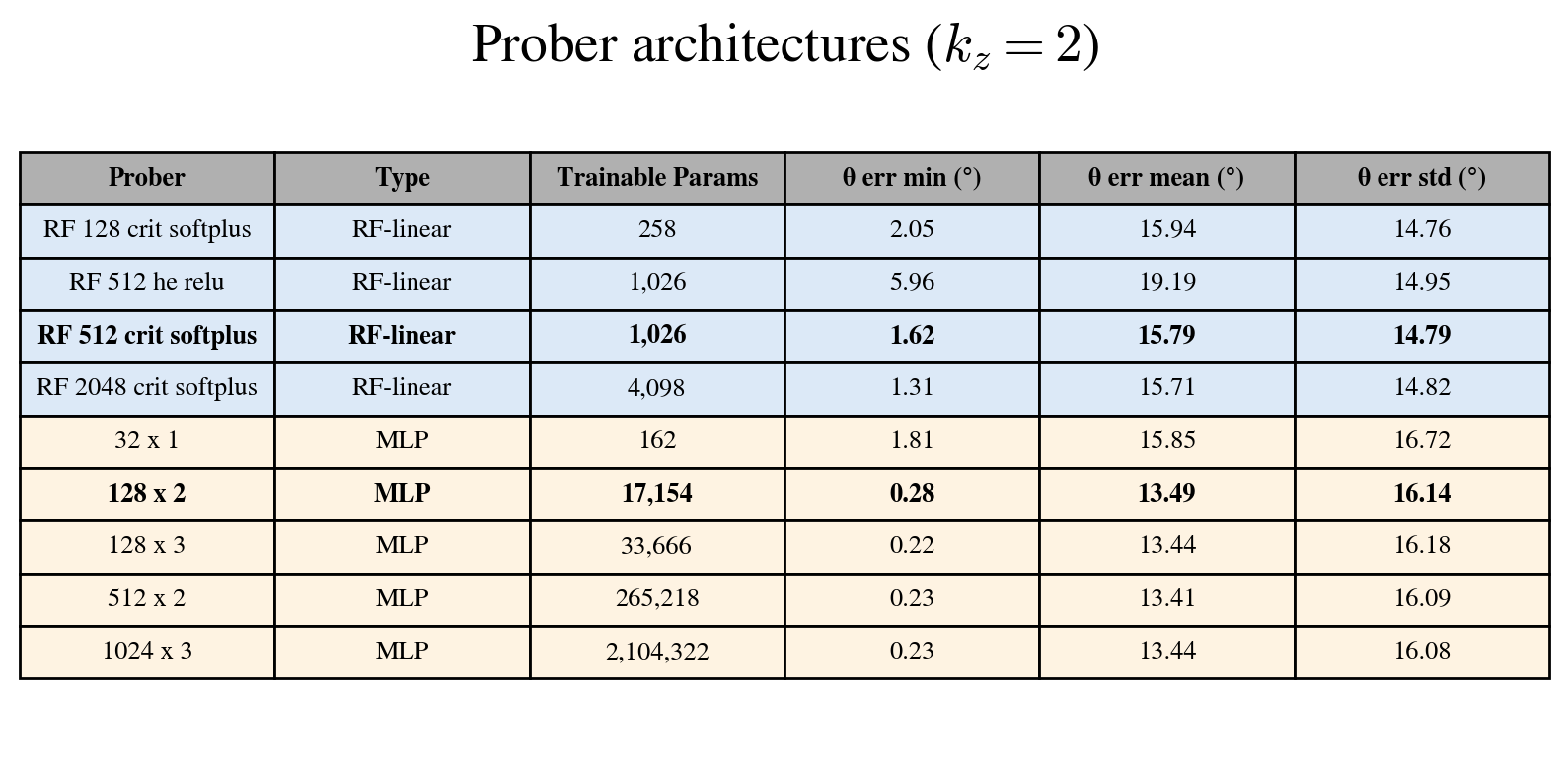}

    \vspace{0.5em}

    \includegraphics[width=0.8\linewidth]{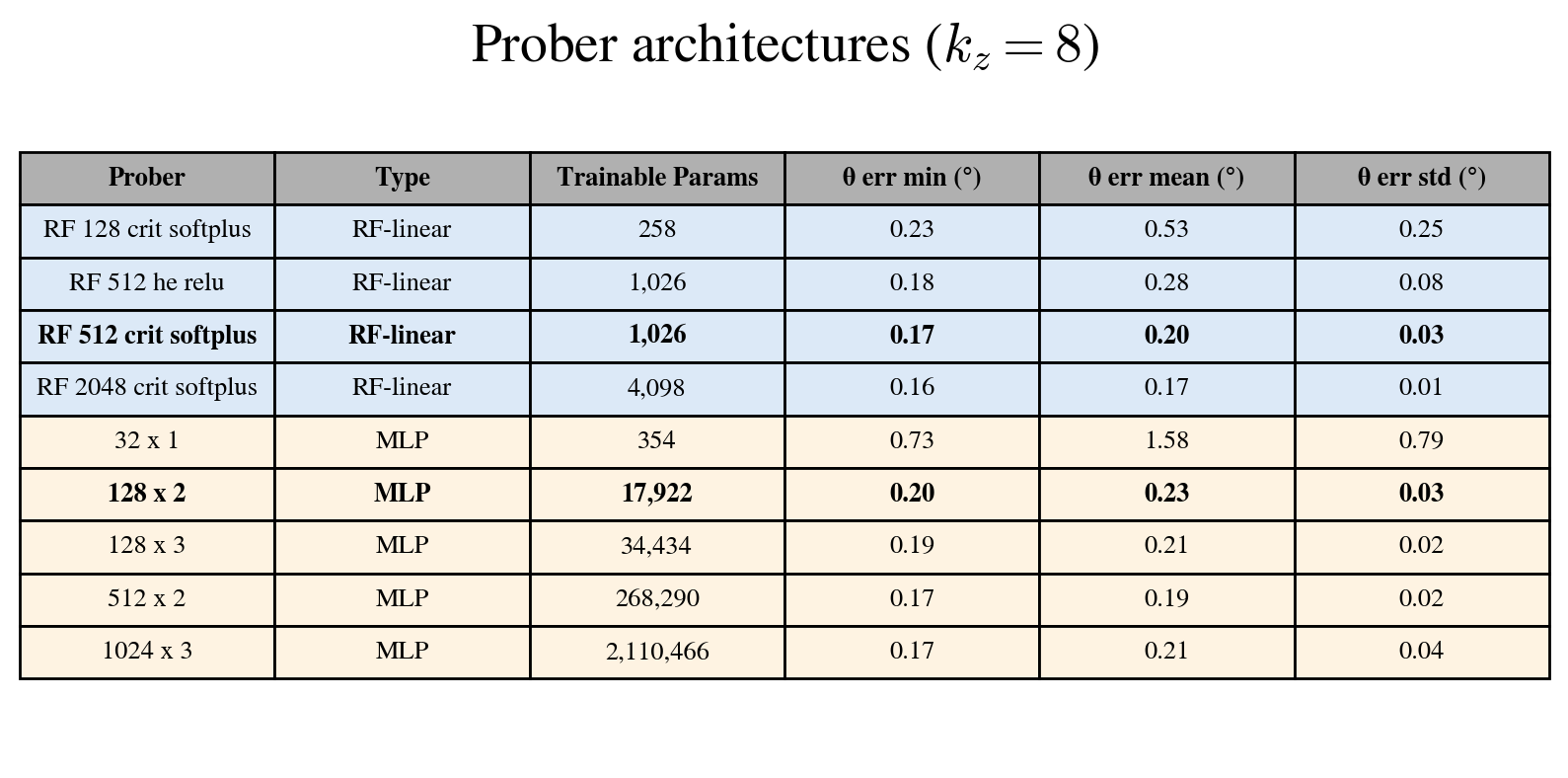}

\caption{
\textbf{Effect of probe architecture.}
We compare random-feature linear probes and MLP probes for \textbf{(A)} the minimal bottleneck, $k_z=2$, and \textbf{(B)} an overparameterized bottleneck, $k_z=8$. Tables report trainable probe parameters and the minimum, mean, and standard deviation of the angle RMSE across 10 independent DySIB training runs. The rows with \textbf{bold} text correspond to the probe architectures used elsewhere throughout the text. In the RF-linear rows, ``RF $m$ crit softplus'' denotes a fixed random expansion to $m$ Softplus features using the critical initialization described in App.~\ref{SI:probers}, followed by a trainable linear readout; ``RF 512 he relu'' denotes the analogous fixed ReLU expansion with He initialization. In the MLP rows, ``$w \times L$'' denotes an MLP with $L$ hidden layers of width $w$. For $k_z=2$, a modest MLP probe improves over the RF-linear probe, consistent with the learned two-dimensional phase-space coordinates being nonlinearly related to the pendulum angle, while larger MLPs give little additional gain. For $k_z=8$, RF-linear probes perform as well as or better than the MLP probes, indicating that the angle is already accessible from simple fixed nonlinear features of the overparameterized latent representation. Across probe architectures, the standard deviation of the angle error is much smaller for $k_z=8$ than for $k_z=2$, consistent with the larger bottleneck reducing trial-to-trial variability.}
    \label{fig:prober_ablation}
\end{figure*}

The probe networks are used only to evaluate the learned representation, but their capacity can affect the numerical value of the reported errors. We therefore compare random-feature linear probes and MLP probes across a range of capacities for both the minimal bottleneck, $k_z=2$, and an overparameterized bottleneck, $k_z=8$ (Fig.~\ref{fig:prober_ablation}). In both cases, the random-feature linear probe performs well, showing that the physical variables are accessible from relatively simple readouts of the learned latent coordinates.

For the minimal bottleneck, $k_z=2$, the learned representation is constrained to encode the full pendulum phase space in only two coordinates. The resulting coordinates recover the correct phase-space structure, but the relation between those coordinates and the canonical variables can be curved and nonlinear. Consistent with this, a modest MLP probe gives a small but meaningful improvement over the random-feature linear probe. Increasing the MLP capacity beyond two hidden layers of width 128 gives little additional improvement, indicating that the benefit comes from allowing a modest nonlinear readout, not from using a large supervised model to reconstruct the physics after the fact.

For the overparameterized bottleneck, $k_z=8$, the random-feature linear probe performs as well as, or better than, the MLP probes. Thus the larger latent space makes the physical variables directly accessible to simple fixed nonlinear features, even though the learned representation remains intrinsically two-dimensional. The $k_z=8$ results also show substantially smaller trial-to-trial variability across probe architectures, consistent with the reduced training variability observed in Fig.~\ref{fig:si_variance}.

\end{document}